\documentclass{article}

\usepackage[utf8]{inputenc}
\usepackage{booktabs} 
\usepackage{graphicx}
\usepackage{grffile}
\usepackage{subfigure}
\usepackage[boxed]{algorithm}
\usepackage{algorithmic}
\usepackage{multicol}
\usepackage[percent]{overpic}
\usepackage{balance}
\usepackage{colortbl}
\usepackage{amssymb} 
\usepackage{natbib}
\usepackage{hyperref}

\definecolor{tabbgd}{rgb}{0.9,0.9,0.9}

\begin{document}

\title{SynGraphy: Succinct Summarisation of Large Networks via Small Synthetic Representative Graphs}

\author{
  Jérôme Kunegis\textsuperscript{1}, Pawan Kumar\textsuperscript{2}, \\
  Jun Sun\textsuperscript{3}, Anna Samoilenko\textsuperscript{2}, and Giuseppe Pirró\textsuperscript{4} \\
  \textsuperscript{1} University of Namur \\
  \textsuperscript{2} University of Koblenz--Landau \\
  \textsuperscript{3} University of Stuttgart \\
  \textsuperscript{4} Institute for High Performance Computing and Networking
}

\maketitle

\begin{abstract}
  We describe SynGraphy, a method for visually summarising the structure
  of large network datasets that works by drawing smaller graphs
  generated to have similar structural properties to the input graphs.
  Visualising complex networks is crucial to understand and make sense
  of networked data and the relationships it represents.  Due to the
  large size of many networks, visualisation is extremely difficult; the
  simple method of \emph{drawing} large networks like those of Facebook
  or Twitter leads to graphics that convey little or no information.
  While modern graph layout algorithms can scale computationally to
  large networks, their output tends to a common \emph{hairball} look,
  which makes it difficult to even distinguish different graphs from
  each other.  Graph sampling and graph coarsening techniques partially
  address these limitations but they are only able to preserve a subset
  of the properties of the original graphs.  In this paper we take the
  problem of visualising large graphs from a novel perspective: we leave
  the original graph's nodes and edges behind, and instead summarise its
  properties such as the clustering coefficient and bipartivity by
  generating a completely new graph whose structural properties match
  that of the original graph.  To verify the utility of this approach as
  compared to other graph visualisation algorithms, we perform an
  experimental evaluation in which we repeatedly asked experimental
  subjects (professionals in graph mining and related areas) to
  determine which of two given graphs has a given structural property
  and then assess which visualisation algorithm helped in identifying
  the correct answer.  Our summarisation approach SynGraphy compares
  favourably to other techniques on a variety of networks.
\end{abstract}

\section{Introduction}
\label{sec:intro}
Network data is nowadays ubiquitous; social networks, protein networks
and the Web are just a few examples of its spread.  If network 
analysis is a crucial challenge to gain insights from such wealth of 
data, network visualisation represents an equally important aspect, 
especially because of its immediate usefulness to both network analysts and 
non-expert users; displaying (portions of) a network or structural properties 
(e.g.\ the degree distribution) makes more accessible the specific task 
at hand.
The visualisation of small graphs poses no major problem; it can be
achieved by graph drawing, i.e., placing each node of a graph at a
specific position on the plane, and drawing edges as lines.  The task
then becomes to find suitable coordinates of nodes such that the drawing
has judicious properties for the user \citep{purchase2005}, including the
fact that connected nodes are 
near to each other, lines should not cross, points and
lines should be uniformly distributed over the drawing area, and symmetries 
should be highlighted.
\begin{figure}[!t]
  \centering
  \begin{overpic}[width=0.20\columnwidth]{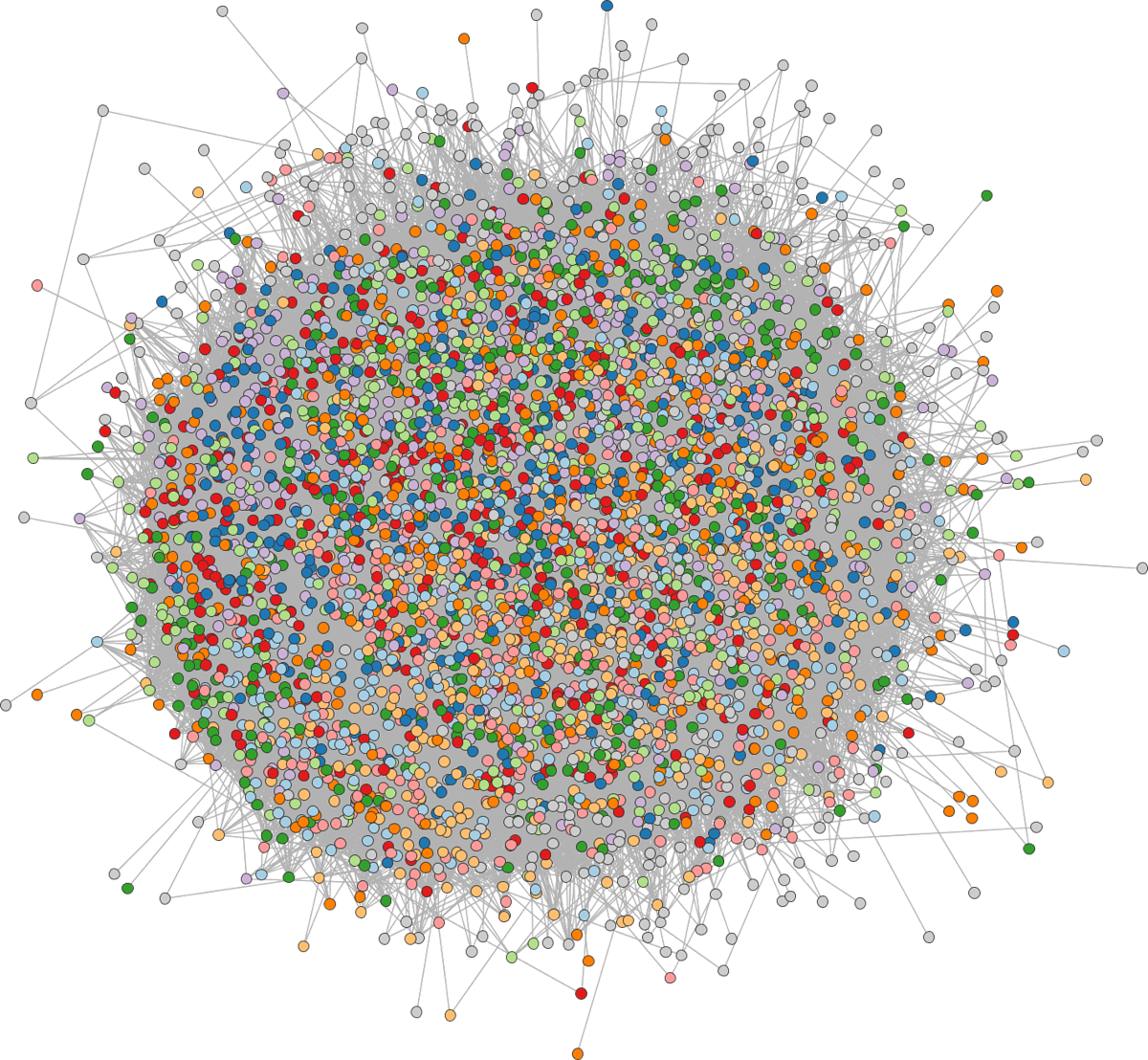}
    \put (5,10) {\textsf{\textbf{A}}}
  \end{overpic}
  \begin{overpic}[width=0.20\columnwidth]{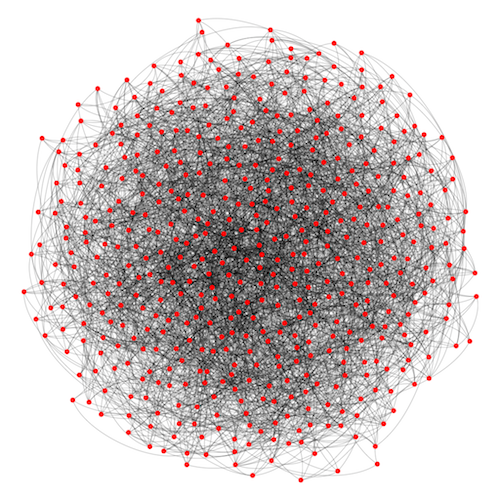} 
    \put (5,10) {\textsf{\textbf{B}}}
  \end{overpic}
  \begin{overpic}[width=0.20\columnwidth]{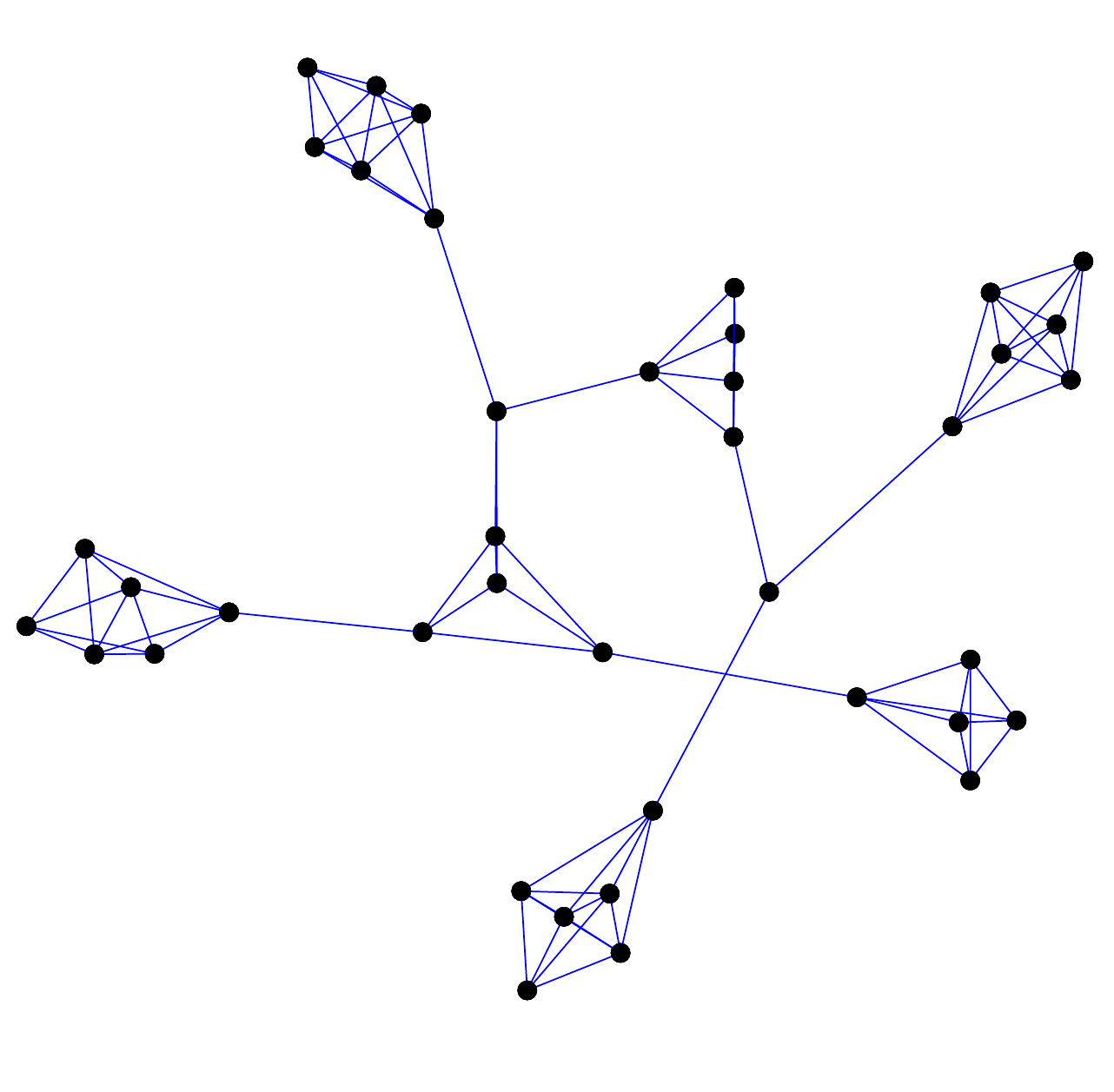}
    \put (5,10) {\textsf{\textbf{C}}}
  \end{overpic}
  \begin{overpic}[width=0.20\columnwidth]{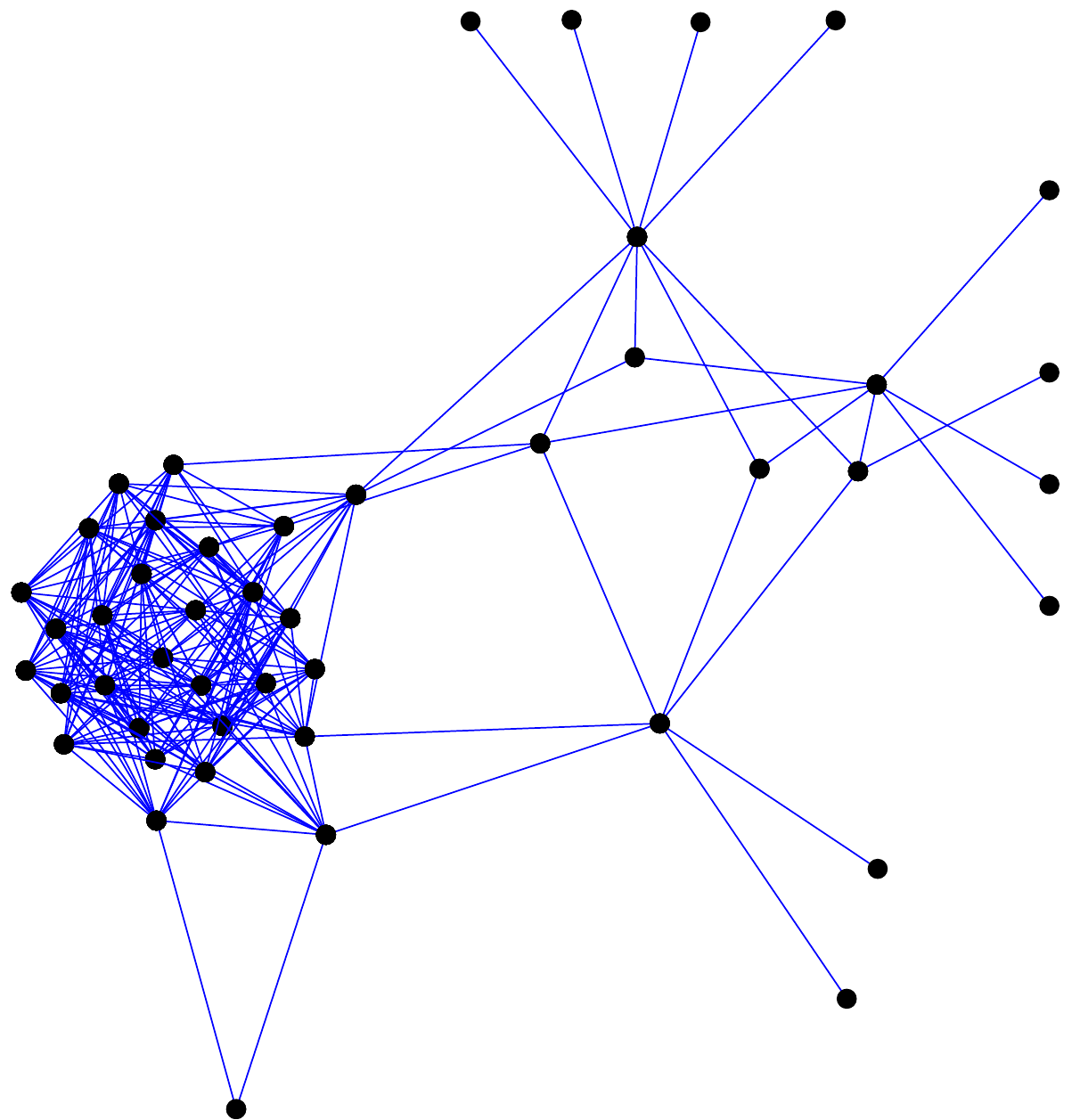}
    \put (5,10) {\textsf{\textbf{D}}}
  \end{overpic}
  \caption{
    Comparing usual \emph{hairball}-like graph drawings
    \textsf{\textbf{(A--B)}}
    with the SynGraphy graph visualisation method introduced in
    this paper \textsf{\textbf{(C--D)}}.
    \textmd{
    \textsf{\textbf{(A)}} a social network \citep{hairball-gallery-1}, 
    \textsf{\textbf{(B)}} a block chain network \citep{hairball-gallery-3},
    \textsf{\textbf{(C)}} an email network \citep{konect:klimt04},
    \textsf{\textbf{(D)}} a loan network \citep{konect:prosper-loans}. 
    The goal of this paper is to avoid the
    ``hairball'' effect and make graph-theoretical features of the
    network apparent. 
    }
    \label{fig:hairball-gallery}
  }
\end{figure}
%
%
%
%
%
%
In data mining applications such as those arising on the web however,
networks are so large that graph drawing is highly problematic, for at least 
two aspects. First, the computational cost associated to a certain
graph drawing method which models properties like those listed above is usually high. 
Second, even if one comes up with a suitable algorithm, 
the drawing of very large graphs inherently leads to
the ``hairball effect'', i.e., the fact that all graphs look the same when 
drawn~-- see the article by Schulz and Hurter (\citeyear{schulz2013grooming}) for a comprehensive overview about 
the topic.  Examples 
of hairball-like plots are shown in
Figure~\ref{fig:hairball-gallery}: even if the networks come from different 
domains, it is difficult to discern any of their structural properties.

To avoid the hairball effect, other
methods are typically used, such as drawing only a part of the network
(graph sampling), or aggregating its nodes into groups to show its main 
overall structural features (graph coarsening).  These methods however lead 
to the problem of representativity, i.e., the sampled or coarsened graphs have 
structural properties differing from the original graphs.  Some methods have 
been developed to bias graph sampling and coarsening algorithms to 
preserve certain graph properties \citep{cem2013}. However, these approaches are 
not focused on taking into account \textit{specific} user's information needs; 
they rely on the layout itself as a means to satisfy all information needs. 

\vspace{0.5cm}
\noindent
\textbf{Contributions.}
In this paper we investigate the problem of 
visualising large graphs from the perspective of graph summarisation.  Instead of applying 
sampling or coarsening, \emph{we devise an algorithm that generates  
  synthetic graphs whose properties match that of the original graphs}. The 
crucial aspect is that the
syntactic graphs generated with our algorithm do not preserve 
any nodes or group of nodes from the original graphs; instead,
they preserve the original structural and graph-theoretical properties.  This 
allows us to reduce the problem of visualising large networks to the problem of 
drawing smaller, representative graphs instead of the original network.
For a given input network, our algorithm SynGraphy performs the following steps: (i)~it
first measures several (and dynamically selectable) statistical 
properties of a network; (ii)~scales them down to a given, small
predefined number of nodes; (iii)~generates a synthetic   
network with the 
required scaled-down statistical properties; and (iv)~produces a graph 
drawing of the synthetic network using a classical graph drawing algorithm.
The scaling down of statistics is trained on a large set of network
datasets, in order to faithfully distinguish particularly low and high
values of each network statistic. The summarisations of networks 
generated by our algorithm are evaluated in an experiment with
human volunteers that are professionals in network analysis
and related fields, asking them to estimate which of two given graphs
has a higher or lower value of a given statistical property (e.g.,
clustering coefficient).  SynGraphy performed well in comparison 
to plain graph drawing algorithms and graph sampling algorithms.  

The remainder of the paper is structured as follows. 
Section~\ref{sec:background} reviews the areas of graph drawing, 
sampling and network characteristics. Section~\ref{sec:our-approach} 
describes the proposed algorithm. The evaluation of our approach is 
discussed in Section~\ref{sec:evaluation}.  We conclude and sketch future 
work in Section~\ref{sec:conclusion}.

\section{Related Work} 
\label{sec:background}
The problem of graph visualisation is of primary importance in multiple
fields of research, as many types of data can be modelled as
networks.  In order to visualise a graph, the most straightforward -- and
most common -- method is to use graph drawing, i.e., producing a plot in
which each node of the graph is represented as a point, and each edge as a
line connecting a pair of points.  Other, more advanced methods may try to 
show only a subset of a graph, or to summarise its properties in a specific
way.  
The visualisation of network datasets can serve multiple purposes that
can be clearly distinguished, and which lead to different visualisation
methods:
\begin{itemize}
\item \textbf{Exploration} allows to explore nodes and edges
  of a graph.  In the simplest case, a graph drawing allows to
  follow individual nodes and edges, as long as the
  graph is not too large.  If the graph is large, a system of
  navigation may be used, in which one may click, for instance, on nodes
  to explore their neighbourhood;
\item \textbf{Structural properties} -- graph visualisation should
  allow to recognise structural properties of the graph drawn as a
  whole, e.g., whether the graph has a high clustering coefficient or
  not \citep{honeycomb}.  This allows to perform other 
  tasks such as  comparing two networks, or finding an outlier in a given set 
  of networks. 
\end{itemize}
In this work the focus is on recognising structural properties: the 
goal is to allow both users and network analysts to recognise such
properties in the graph visualisation, thus leading to the task of
graph summarisation -- 
the exploration of nodes and edges of 
the graph is then an orthogonal goal to ours.

\subsection{Graph Drawing}
Graph drawing, as opposed to graph visualisation, specifically refers to
the task of laying out nodes and edges of a graph on a planar
surface in order to produce a figure that shows the input
graph as a whole. Graph drawing has also been called graph layouting and
graph embedding. 
In general, graph drawing algorithms are based on optimising a certain cost
or utility function, which models the quality of the graph drawing.
Examples of features that can be incorporated are: drawing nodes
near to their neighbours, avoiding crossing of edges, and avoiding multiple
nodes at the same coordinates.  A subset of such algorithms performs a
force-directed layout, i.e., it simulates a physical force between nodes
based on the physical model of springs connected the nodes.  In very simple cases,
these can be formulated as the eigenvectors of certain graph-characteristic 
matrices, in particular of the Laplacian matrix \citep{b405}. 
A well-known graph layout algorithm that is widespread in practice is the
algorithm of Fruchterman and Reingold (\citeyear{b870}), which bases its
physical model on placing nodes near to its neighbours, but also on
not placing nodes too near to each other -- this makes the problem
nonlinear and thus it cannot be expressed as eigenvectors of a matrix,
but instead can be solved efficiently by direct optimisation algorithms, even if a
global minimum of the energy function cannot always be found. 

The obvious problem with pure graph drawing methods is that they don't
scale visually -- if the number of nodes is too high, the resulting drawing will
contain so many dots and lines that it will look like a single big patch
of dots, without any recognisable network structure.  This is the type
of drawing that is usually called a ``hairball'', and which is illustrated in 
Figure \ref{fig:hairball-gallery} using examples from the literature. 

Another class of graph drawing algorithm approaches the problem from
the point of view of wanting to draw the network such that certain
particular properties of the network are highlighted, and thus choose to
optimise the layout such that the given properties are made visible.  An
example from this class of algorithms is the hive plot \citep{hive-plots}, in
which nodes of the given graph are assigned to one of three or more
axes, which may be divided into segments. Nodes are ordered on a segment
based on properties such as connectivity, density, centrality or
quantitative annotations. The user then is free to choose whatever rules
fit their data and visualisation requirements. Edges are drawn as
curves.  As we will see, the SynGraphy method presented in this paper
has in common with hive plots that it makes it possible to assess
network structure because both are based on network properties, not on
aesthetic layout, making visualisations of two networks directly
comparable.

\subsection{Graph Summarisation}
The inadequacy of drawing a complete graph when its size is
too large leads to several methods in which the size of the input
graph is reduced to obtain a smaller graph, which can then be drawn using
any of the methods mentioned in the previous section.  
Such methods can
be roughly divided into two types: (i)~graph sampling techniques which
choose a subset of the given graph's nodes and edges, and (ii)~graph
coarsening techniques that aggregate the graph's nodes into new
nodes.  
These methods can be interpreted as graph summarisation:
Automatic summarisation
refers to the generation of shorter documents that summarise a given
document, reducing it to the most important parts.  
Although the term \emph{summarisation} has recently been applied to graphs, for
instance by Liu and colleagues (\citeyear{graph-summarization}), the term is, as
of 2020, not common in the field. 

\noindent
\textbf{Graph sampling.}
Graph sampling or graph sparsification refers to reducing a graph to a
subset of its nodes and edges, with the goal of preserving 
certain graph properties.  Simple
graph sampling techniques perform a random sampling of nodes and edges, 
where each node or each edge is kept with a given
probability \citep{hu:graph-sampling}.  In such methods, the expected properties of the
resulting network can sometimes be derived in closed form.  As an example,
an edge sampling procedure that keeps each edge with probability $p$
will lead to a graph with clustering coefficient $pc$, where $c$ is the
original graph's clustering coefficient.  On the other hand, certain
graph sampling methods do not choose nodes and edges randomly, but use
heuristics. For instance, an approach to filter a
graph by removing edges in order of increasing betweenness
centrality is described by Jia and colleagues (\citeyear{onVisualSocial}).  
The method preserves certain properties such as the connectivity,
but there are still some features in the resulting graph
that are hidden and cannot be identified for the entire graph.     

\noindent
\textbf{Graph coarsening.}
Graph coarsening consists in aggregating multiple nodes of the input
graph into a single node of the output graph, essentially performing
vertex identification or edge contraction \citep{openord}.  As a result, high-level
structural features such as connections between large communities
are preserved, while information at the node level such as the
clustering coefficient is usually lost. 
More examples (and drawings) of graph coarsening methods are given by
Nocaj and colleagues (\citeyear{nocaj2015}). 

\subsection{Graph Characteristics}
\label{sec:graph-characteristics}
In order to generate a graph with similar characteristics to an
input graph, the characteristics of the latter need to be numerically 
quantified.  This is achieved through the use of \emph{graph
  statistics}, i.e., numerical measures of a graph.  Graph statistics
comprise the important class of subgraph counts, which count the number
of occurrences of a particular small graph as a subgraph.  For instance,
the number of triangles is an often used graph statistic.  The number
of nodes and edges of a graph are subgraph counts too, corresponding to
the subgraph consisting of a single node and a single edge
respectively.  Other graph statistics may be derived from subgraph
counts, such as the clustering coefficient, which can be expressed as
$3t/s$, where $t$ is the number of triangles and $s$ is the number of
wedges, while other graph statistics are wholly unrelated to subgraph
counts (e.g., the diameter). 
The graph statistics useful for the purposes of our algorithm, and described 
below, are summarised in Table~\ref{tab:notation}. The input graph, as well 
as all other graphs used in this paper are undirected, without parallel edges, 
and loopless. 
\begin{table}
  \caption{
    \label{tab:notation}
    The notation used in this paper.  
    \textmd{
    As a general rule, symbols without
    an apostrophe denote the potentially large input graph, while symbols with an
    apostrophe denote the small output graph to be drawn. 
    }
  }
  \centering
  \begin{tabular}{ c l }
    \textbf{Notation} & \textbf{Meaning} \\
    \hline
    $G=(V,E)$ & The input graph (potentially large) \\
    \rowcolor{tabbgd}$G'=(V',E')$ & The output graph (to be drawn, small) \\
    \hline
    $n$ & Number of nodes \\
    \rowcolor{tabbgd}$m$ & Number of edges \\
    $s$ & Number of wedges \\
    \rowcolor{tabbgd}$z$ & Number of claws \\
    $x$ & Number of crosses \\
    \rowcolor{tabbgd}$t$ & Number of triangles \\
    $q$ & Number of squares \\
    \rowcolor{tabbgd}$d$ & Average degree \\
    $c$ & Clustering coefficient \\
    \rowcolor{tabbgd}$y$ & 4-Clustering coefficient \\
    $b$ & Bipartivity \\
    \rowcolor{tabbgd}$\delta$ & Diameter \\
    $\rho$ & Degree assortativity \\
    \hline
    \rowcolor{tabbgd}$\square'$ & Scaled-down value of any statistic $\square$ \\
    \hline
  \end{tabular}
\end{table}  


\noindent \textbf{Network size and volume ($n, m$).}
The number of nodes and edges in a network. 
\vspace{0.1cm}

\noindent \textbf{Number of wedges ($s$).}
A wedge is a subgraph consisting of 
a central node connected to two other nodes.  Equivalently, wedges are 
2-stars or 2-paths, and their count can be expressed as
\begin{equation}
  s = \sum_{u\in V} \frac{d(u)(d(u)-1)}{2},
\end{equation}
where $d(u)$ is the degree of node $u$. 
\vspace{0.1cm}

\noindent \textbf{Number of claws ($z$).}
A claw is a subgraph equivalent to a 
3-star. 
\vspace{0.1cm}

\noindent \textbf{Number of triangles ($t$).}
A triangle is a subgraph consisting 
of three mutually connected nodes. 
\vspace{0.1cm}

\noindent \textbf{Number of squares ($q$).}
A square is a group of four nodes 
connected to each other in the shape
of a square, i.e., a cycle of length four.
\vspace{0.1cm}

\noindent \textbf{Clustering coefficient ($c$).}
The clustering coefficient 
represents the presence of
a heightened number of triangles in the network \citep{complexNwk}, and
can be expressed as
\begin{equation}
  c = \frac{3 t}{s}.
\end{equation}
The clustering coefficient can take a value between zero and one,
where the value one denotes that all possible triangles are present, and zero
denotes a triangle-free network.  
\vspace{0.1cm}

\noindent \textbf{4-Clustering coefficient ($y$).}
The 4-clustering coefficient $y$ 
is analogous to the clustering
coefficient, but measures the probability that a randomly chosen path of
length three is closed.  It equals
\begin{equation}
  y = \frac {4 q} {P_3},
\end{equation}
where $P_3$ is the number of paths of length three.
The 4-clustering coefficient complements the clustering coefficient, and
may be used in bipartite or near-bipartite networks \citep{b795}. 
\vspace{0.1cm}

\noindent \textbf{Bipartivity ($b$).}
We use as a measure of bipartivity the ratio of the smallest and largest
eigenvalue of the graph's adjacency matrix $\mathbf A$ \citep{kunegis:bipartivity}:
\begin{equation}
  b = \left| \frac {\lambda_{\min}[\mathbf A]} {\lambda_{\max}[\mathbf A]} \right|
\end{equation}
\vspace{0.1cm}

\noindent \textbf{Average degree ($d$).}
The average degree measures how many edges are connected to a node on
average. The average degree is sometimes called the density.  It equals $d=2m/n$. 
\vspace{0.1cm}

\noindent \textbf{Diameter ($\delta$).}
The diameter of a network is the longest shortest distance between any
two nodes. A distance is measured by the number of edges needed to reach
one node from another.
\vspace{0.1cm}

\noindent \textbf{Degree assortativity ($\rho$).}
In some networks, the nodes with high degree are connected to other
nodes with high degree, while nodes with low degree are connected to
nodes with low degree.  This is measured by the Pearson correlation
coefficient, taken over all pairs of connected nodes \citep{b854}. 

\section{The SynGraphy Approach}
\label{sec:our-approach}
We are now ready to introduce our algorithm. Let $G = (V, E)$ be the large 
input graph that is to be summarised. Our algorithm 
proceeds in four steps:
\begin{enumerate}
\item[(i)] Compute the subgraph counts of $G$
\item[(ii)] Scale down the subgraph counts
\item[(iii)] Generate $G'$ with the scaled down subgraph counts
\item[(iv)] Generate a drawing of $G'$
\end{enumerate}
Steps (i--iii) are
described in the following subsections, while Step~(iv) uses the
algorithm by Fruchterman and Reingold (\citeyear{b870}).  For Step (ii), we describe
two variants.  

\subsection{Scaling Down of Graph Statistics}
\begin{figure}
  \centering
  \includegraphics[width=0.50\columnwidth]{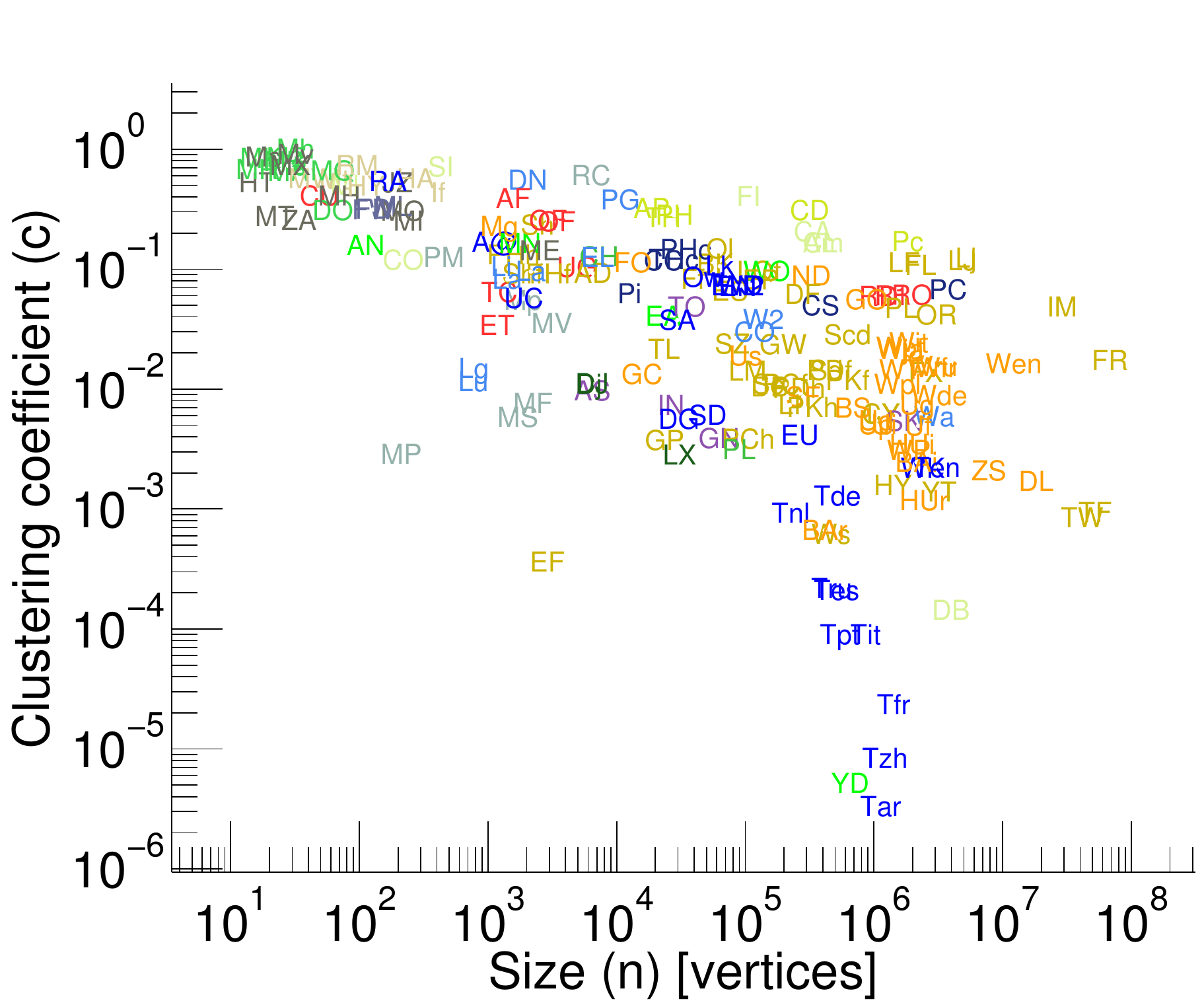}
  \caption{
    The size $n$ versus the clustering coefficient $c$ in real-world
    networks. 
    \textmd{
      This plot shows that the clustering coefficient is not independent
      of size, and gives an approximate relationship of $c \sim
      n^{-0.5}$.  
      The networks are used as provided by the KONECT project
      \citep{konect}. 
    }
    \label{fig:size-clusco}
  }
\end{figure}

The core idea of our algorithm to summarise the input graph $G$ is to generate 
a new graph $G'$ with analogous structural properties to 
those of 
$G$.  Hence, the focus is on providing a meaningful rendering of $G$'s 
structural properties instead of $G$'s aesthetic. 
The size of $G'$, i.e., the 
number of nodes $n'$, can be chosen arbitrarily.  For the sake of presentation, 
in what follows we set it to the value $n'=80$.  As for the other statistics, 
their new values in the output graph $G'$ need to be computed before $G'$ 
can be generated. Certain graph statistics, such as the average degree $d$ 
and clustering coefficient $c$ are in principle scale-invariant, and could
be simply carried over from $G$ without modification, i.e., it is
possible to use $d'=d$ and $c'=c$.  Other statistics such as the number
of triangles clearly scale with $n$ and their values have to be
adjusted.  In fact, the values of $c$ and $d$ may have to be adjusted
too, as illustrated in Figure~\ref{fig:size-clusco}.  In this light, we investigate two methods 
for determining the new values of these graph statistics:
\begin{itemize}
\item (\textsf{\textbf{NO}}): normalising all statistics empirically;
\item (\textsf{\textbf{SI}}): use of statistics that by construction are
  independent of size.
\end{itemize}

\noindent \textbf{Empirical Normal Distribution (\textsf{\textbf{NO}}).}
One possible assumption is that when normalised in certain ways, certain
network statistics are size-independent.
An example for this is the clustering coefficient.  However,
as we can derive empirically from actual network datasets, this is not
entirely correct:  the clustering coefficient does correlate with the
size of the network, when measured in real-world networks, as shown in
Figure \ref{fig:size-clusco}.  This observation leads to the following method
for scaling down graph statistics:  fit a distribution to
the statistics of a set of known networks, and then read out the most
likely values of the requested graph statistics, given the target size
$n'$. 
Given a set of real-world networks and a set of statistics, we compute
the values of the statistics for all networks, and then fit a
multivariate normal distribution to the data.  
The multivariate normal density is a generalisation of the
univariate normal density to $k \geq 2$ dimensions \citep{mvnd}. The
$k$-dimensional normal density of a $k$-dimensional variable $\mathbf x$
takes the form:
\begin{equation}
  f(\mathbf x) = 
  \frac 1 {\sqrt{(2\pi)^k |\mathbf\Sigma|}}
  \exp\left\{-\frac 12 (\mathbf x- \mathbf\mu )^{\mathrm T}\mathbf \Sigma^{-1}(\mathbf x-\mathbf\mu)\right\},
\end{equation}
where the $k$-vector $\mathbf\mu$ is the mean of the
distribution, the $k \times k$ matrix $\mathbf\Sigma$ the
covariance matrix, and $|\mathbf\Sigma|$ its determinant. 
The conditional probability distribution under a fixed $n=n'$ is itself
a multivariate normal distribution whose mean is given by 
\begin{equation}
  \mathbf\mu' = 
  \mathbf\mu_* - \frac{\mathbf\mu_n - n'}{\mathbf\Sigma_{n,n}}
  \mathbf\Sigma_{*,n}^{\phantom{-1}} ,
\end{equation}
where the index $\square_n$ denotes a restriction of the given quantity to the
value corresponding to the size statistic $n$, and $\square_*$ denotes a
restriction to all other columns and rows. 
Thus, for given statistic $\alpha$, the new value for the smaller graph is given by
\begin{equation}
  \alpha' = \alpha - (\mu_n - n')\frac {\mathbf \Sigma_{\alpha,n}}{\mathbf \Sigma_{n,n}} 
\end{equation}

\noindent \textbf{Size-Independent Statistics (\textsf{\textbf{SI}}).}
Under the assumption that most subgraph counts can be expected to have values growing with the
number of nodes in the network, the number of edges $m$ for instance
can be written as
\begin{equation}
  m = \frac d 2 n,
\end{equation}
where $d$ is the average degree in the network.  In a social network for
instance, $d$ represents the average number of friends, and we can make
the reasonable assumption that $d$ is independent from the network size, 
i.e., people don't have more friends because the world is larger.  Note that
this assumption is not necessarily true, and the opposite case is taken into 
account in the previous paragraph.  Assuming that $d$ is constant, it follows 
that we can use $d$ as a size-independent statistic, i.e., we can compute $d$
for the input graph, and then use $d'=d$. 

Similarly, other subgraph counts can be derived to grow as powers of the
network's size, given an underlying graph model.  In fact, this is true
for whole classes of network models; in the Erdős--Rényi model for
instance, the expected value of any subgraph count is a polynomial of
the graph size \citep{gp6}.  For instance, the number of triangles $t$ in
Erdős--Rényi graphs grows as the cube of the number of vertices. 

For each subgraph count $\alpha$, we perform the following procedure:
We derive the expected value of $\alpha$ under suitable conditions, calling it $\hat
\alpha$; and then derive from it the normalised statistic $\alpha_{\mathrm 
  n}^{\phantom{x}} = \alpha /
\hat \alpha$.  Finally, we use $\alpha_{\mathrm n}' = \alpha_{\mathrm
  n}^{\phantom{x}}$, giving the new value of the plain subgraph count:
\begin{equation}
  \alpha' =  \frac{\hat \alpha'}{\hat \alpha\phantom{'}} \alpha 
\end{equation}

We now apply this method to individual subgraph count statistics.  For
each subgraph count statistic $\alpha$, we derive the expected value
$\hat \alpha$ and the scaled down subgraph count statistic $\alpha'$.
\vspace{0.1cm}

\noindent \underline{Number of edges ($m$)}:
Given a fixed average degree $d' = d = 2m/n$, we get
\begin{eqnarray} 
  \hat m = \frac d 2 n, \qquad
  m'     = \frac {n'} {n\phantom{'}} m. 
\end{eqnarray}

\noindent \underline{Number of wedges ($s$)}:
Given an average degree $d$ and an equal distribution of degrees, the
expected number of wedges can be computed by noting that the number of
wedges $s(u)$ centred at each node $u$ is given by
${d(u) \choose 2}$.  
\begin{eqnarray}
  \hat s = n {d \choose 2}, \qquad
  s'     = \frac {n'} {n\phantom{'}} s
\end{eqnarray}

\noindent \underline{Number of claws and crosses ($z, x$)}:
Claws and crosses are 3-stars and 4-stars and as such, their expected
values are analogous to those for 2-stars: 
\begin{eqnarray}
  \hat z = n {d \choose 3}, \qquad
  \hat x = n {d \choose 4}, \qquad
  z'     = \frac {n'} {n\phantom{'}} z, \qquad
  x'     = \frac {n'} {n\phantom{'}} x
\end{eqnarray}

\noindent \underline{Number of triangles ($t$)}:
For the number of triangles, we assume a constant clustering coefficient
$c$.  The clustering coefficient can be expressed by definition as
$c=3t/s$.  This gives
\begin{eqnarray}
  \hat t = \frac c 3 \hat s = \frac c 3 {d \choose 2} n, \qquad
  t' = \frac {n'} {n\phantom{'}} t.
\end{eqnarray}

\noindent \underline{Number of squares ($q$)}:
This case is defined analogously to the number of triangles. 
\begin{eqnarray}
  \hat q = \frac y 4 m (d-1)^2 = \frac y 4 \frac d 2 (d-1)^2 n, \qquad
  q'     = \frac {n'} {n\phantom{'}} q
\end{eqnarray}

\subsection{Graph Generation}
\label{sec:graph-generation}
As previously mentioned, graphs are generated by taking the fixed size of 
the network ($n'=80$), and corresponding scaled down values of six 
statistics including the count of edges, wedges, claws, crosses, triangles, and
squares as obtained by applying the two methods (i.e., \textsf{\textbf{NO}} and 
\textsf{\textbf{SI}}) discussed above. The layout is drawn using the force-directed 
algorithm of Fruchterman and Reingold (\citeyear{b870}).

The general form of our graph
generation algorithm applying to any set of network statistics is given
in Algorithm~\ref{alg:gg}. The algorithm works by taking as a starting point 
an Erdős--Rényi graph with the correct number of nodes and edges, and 
then modifying the graph
iteratively until the resulting graph is near enough to the target.  At
each step of the iteration, we need to consider a certain number of
possible changes in the graph, and choose the change which leads to the
lowest error measure.  In order to compute the changes in the statistics
efficiently, individual changes that are considered should be small,
such that the changes in the statistic values can be easily
computed. The smallest change that we can make in a graph (without
changing the node set) is to add or remove an edge.  In fact, the change
in subgraph count statistics for the addition and removal of an edge can
be expressed in terms of the immediate properties of the two involved
nodes.  Furthermore, in order to optimise the algorithm,
the computation of the changes in statistics over all changes considered
in one step makes use of common subexpression.  Thus, we consider
at each step the additions and removals of edges connected to one single
node.  As we show in the next section, this leads to efficient
expressions for the change in the subgraph count statistics. 
We also note that the described algorithm, while performing the changes
with best reduction in error at each step, is not a pure greedy algorithm; 
indeed, the algorithm continues to iterate in the case that the error $e$ cannot be further
reduced by any single edge switch.  

\begin{algorithm}
  \caption{
    \label{alg:gg}
    The small graph generation algorithm in non-vectorised form. 
  }
  \begin{algorithmic}
    \REQUIRE a size $n'$
    \REQUIRE a set of statistic labels $\mathcal S = \{m,s,z,x,t,q\}$ and
    their values $\mathbf x$ for the input graph
    \REQUIRE a convergence parameter $\epsilon > 0$
    \ENSURE a graph $G'=(V',E')$
    \\\qquad \\
    \STATE $G' = \textrm{ErdősRényi}(n', m' / {n' \choose 2})$
    \FORALL {$S \in \mathcal S$}
    \STATE $\mathbf y^S = S(G')$
    \ENDFOR
    \REPEAT 
    \STATE Choose a node $u \in V'$ at random
    \FORALL {$S \in \mathcal S$}
    \FORALL {$w \in V' \setminus \{u\}$}
    \STATE {$\Delta^S_w = S(G' \pm \{u,w\}) - S(G')$}
    \ENDFOR
    \ENDFOR
    \STATE $v = \mathrm{argmin}_{w\in V' \setminus \{u\}} \sum_{S \in
      \mathcal S} ((\mathbf y^S + \Delta^S_w - \mathbf x^S) / \mathbf x^S)^2$
    \STATE{$G' = G' \pm \{u,v\}$}
    \FORALL {$S \in \mathcal S$}
    \STATE {$\mathbf y^S = \mathbf y^S + \Delta^S_v$}
    \ENDFOR
    \STATE $e = \sum_{S \in \mathcal S} ((\mathbf y^S - \mathbf x^S) / \mathbf x^S)^2$
    \UNTIL {$e$ has not achieved a new minimum value in the last
      $(-n' \log \epsilon)$ iterations}
  \end{algorithmic}
\end{algorithm}

The function $\textrm{ErdősRényi}(n', p)$ generates an Erdős--Rényi graph with
$n'$ vertices and individual edge probability $p$. 
$G' \pm \{u,v\}$ denotes the graph $G'$ in which the state of the edge
$\{u,v\}$ has been switched, i.e.\ removed or added depending on
whether $\{u,v\}$ is present or not. 
The convergence parameter $\epsilon$ ensures that the expected ratio of
nodes $u$ that were not visited since the last new minimum value
$e$ was found equals $\epsilon$.  In all experiments, we use a value of
$\epsilon = 0.01$. 

Algorithm~\ref{alg:gg} requires to compute the difference in statistics
between the current graph $G'$ and the graph $G'$ in which one edge is
added or removed:
\begin{equation}
  \Delta^S_w = S(G' \pm \{u,w\}) - S(G')
\end{equation}
In order to speed up the algorithm, this calculation is performed in a 
vectorised way.  The existence of closed-form expressions
for $\Delta^S_w$ decides whether a particular statistic can be used in our
algorithm. 
Since $\Delta^S_w$ must be computed at each step for all nodes $w\in V$
(except for $w=v$), we give vectorised expressions that
give a vector $\Delta^S$ containing the value $\Delta^S_w$ for all $w\in
V$.  The individual value computed for $w=v$ is then simply ignored.  
During the run of the algorithm, the graph $G'$ is always represented by
its symmetric adjacency matrix $\mathbf A \in \{0,1\}^{n \times n}$.

In what follows we provide expressions for the vectors $\Delta^S$ 
measuring the
change in statistic $S$ expressed as functions of the graph's adjacency matrix $\mathbf A$, for each
statistic $S \in \{m, s, z, x, t, q\}$. 
These expressions make use of the degree vector $\mathbf d$, which is 
updated along with the matrix $\mathbf A$. 
$\mathbf u \circ \mathbf v$ denotes the entry-wise product between
two vectors $\mathbf u$ and $\mathbf v$, and $\mathbf A_{:u}$, the
$u$\textsuperscript{th} column of $\mathbf A$. 
\vspace{0.1cm}

\noindent
\underline{Number of edges ($m$)}:
The number of edges will always increase or decrease by one, depending
on the previous state of the edge $\{u,w\}$.
\begin{equation}
  \Delta^m = - 2 \mathbf A_{:u} + 1
\end{equation}

\noindent
\underline{Number of wedges ($s$)}:
When adding an edge, the number of wedges increases by the sum of the
degrees of the two connected nodes.  When removing an edge, the number
of wedges decreases by the sum of the degrees of the two nodes, minus
two.  
\begin{equation}
  \Delta^s = \Delta^m \circ (\mathbf d + \mathbf d_u) + 2 \mathbf A_{:u}
\end{equation}

\noindent
\underline{Number of claws ($z$)}:
When the two nodes $u$ and $w$ are not connected, the number of added
claws equals $\mathbf d_u (\mathbf d_u-1) + \mathbf d_w (\mathbf
d_w-1)$. When the two nodes are connected, the number of removed wedges
can be computed in the same way, but based on the degrees after the
removal.  
\begin{eqnarray}
  \Delta^z = \frac 12 \Delta^m \circ [(\mathbf d - \mathbf A_{:u})
    \circ (\mathbf d - 1 - \mathbf A_{:u}) \\ 
    \nonumber \qquad \qquad \qquad \qquad \qquad + (-\mathbf A_{:u} + \mathbf
    d_u) \circ (-\mathbf A_{:u} -1 + \mathbf d_u)]
\end{eqnarray}

\noindent
\underline{Number of crosses ($x$)}:
The expression for the number of $k$-stars for higher $k$ follows the
same pattern as for $k=2$ and $k=3$.  Due to the asymmetry between the
addition and the removal of edges, the resulting expressions get
increasingly complex. 
\begin{eqnarray}
  \Delta^x =
  \frac 16 [
    (\mathbf d-1) \circ (\mathbf d-2) \circ (\mathbf A_{:u} \circ (-2
    \mathbf d + 3) + \mathbf d) \\
    \nonumber \qquad \qquad \;+ (\mathbf d_u-1)(\mathbf d_u-2)((3 - 2 \mathbf d_u) \mathbf A_{:u} + \mathbf d_u)
  ]
\end{eqnarray}

\noindent
\underline{Number of triangles ($t$)}:
When adding an edge between two nodes, the number of added triangles
equals the number of common neighbours between the two nodes.  Likewise
when removing an edge, the number of removed triangles equals the
number of common neighbours of the two nodes.  We thus get the following
expression for the change in the number of triangles.
\begin{equation}
  \Delta^t = (\mathbf A \mathbf A_{:u}) \circ \Delta^m
\end{equation}
We thus need to perform one sparse matrix-vector multiplication for each
iteration step. 
\vspace{0.1cm}

\noindent
\underline{Number of squares ($q$)}:
To compute the number of squares added or removed, we count the number
of paths of length three added or removed between $u$ and $w$.  In
principle, this can be achieved by using the expression for $\Delta^t$
and multiplying $\mathbf A_{:u}$ once more by $\mathbf A$.  However,
this will also include the number of paths of length three that include
an edge $\{u,x\}$ or $\{w,x\}$ multiple times, or that include the edge
$\{u,w\}$ if it is present.  Thus, these cases must be subtracted to get
the correct number of squares added or removed.
\begin{equation}
  \Delta^q = (\mathbf A^2 \mathbf A_{:u}) \circ \Delta^m +
  \mathbf A_{:u} \circ (\mathbf d + \mathbf d_u -1 ) 
\end{equation}
We thus need to perform two matrix-vector multiplications in this step. 

\section{Experimental Evaluation}
\label{sec:evaluation}
In this section, we describe the user experiments conducted in order
to evaluate the two proposed graph visualisation methods against
baseline methods, as well as evaluate the runtime of SynGraphy
empirically, and perform an experiment to determine the influence of the
size $n'$ of the generated graphs on the drawings. 

\begin{figure}
  \makebox[\textwidth]{
  \begin{tabular}{cccccc}
    \includegraphics[height=2.4cm]{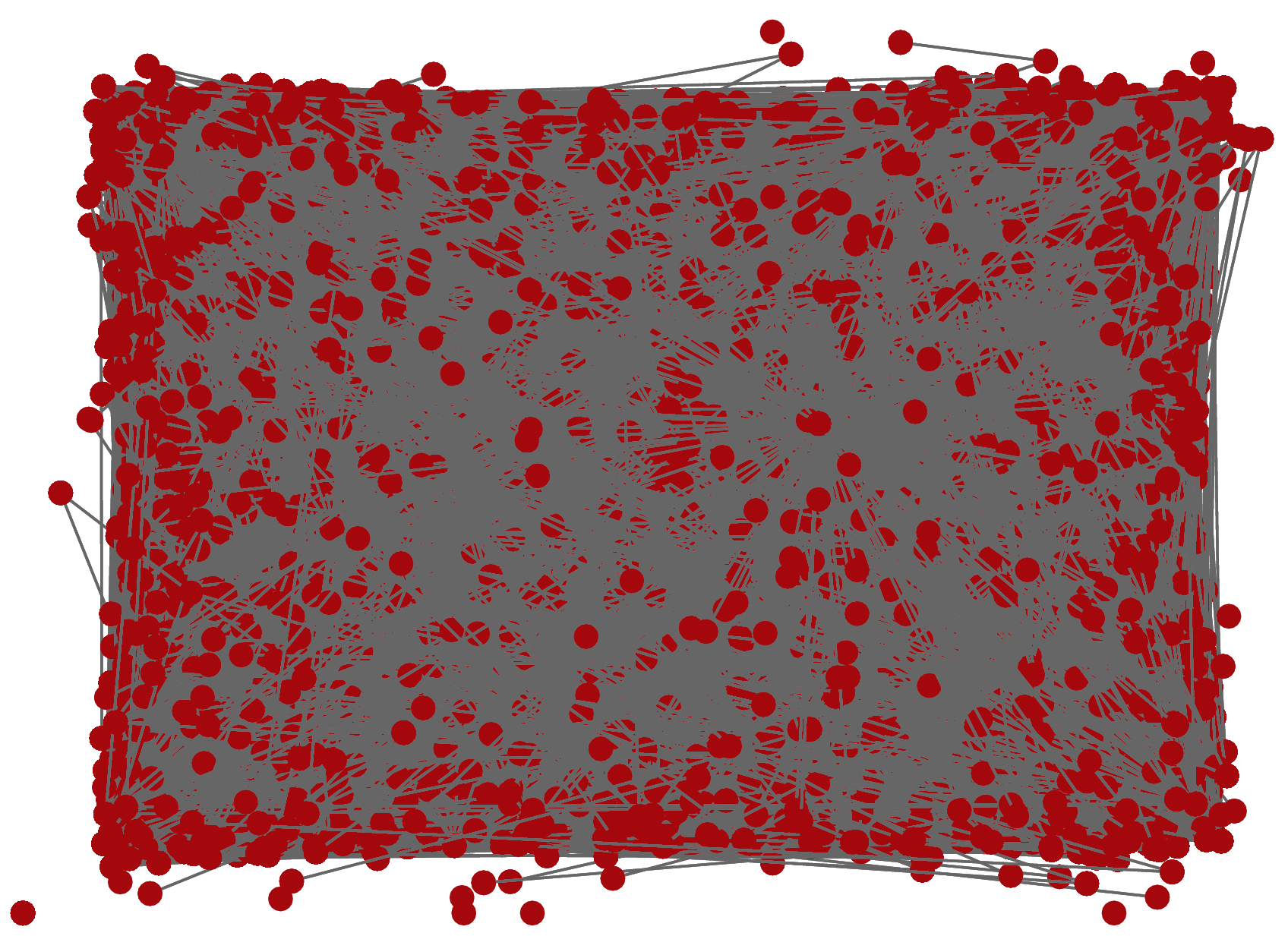} & 
    \includegraphics[height=2.4cm]{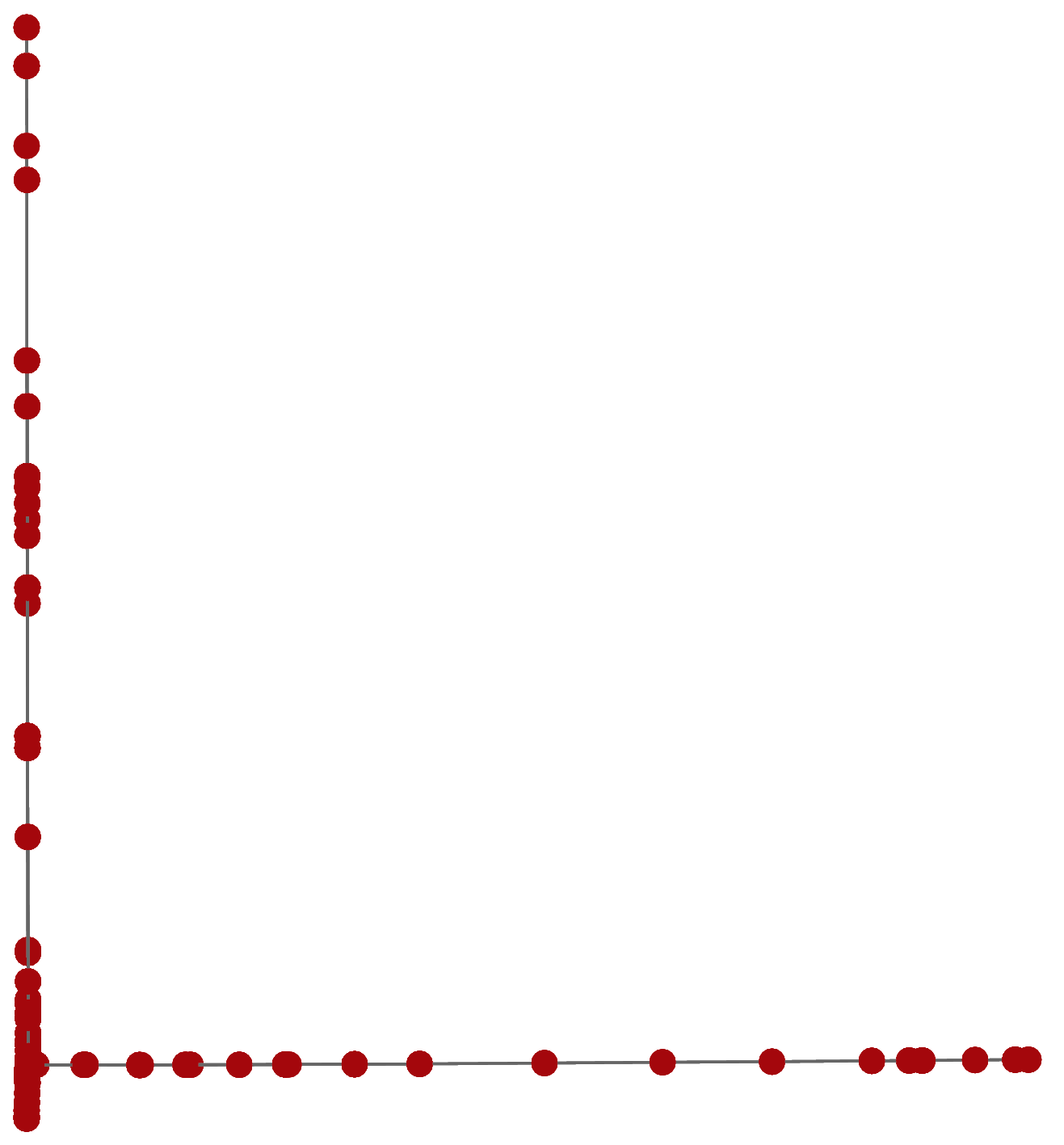} &
    \includegraphics[height=2.4cm]{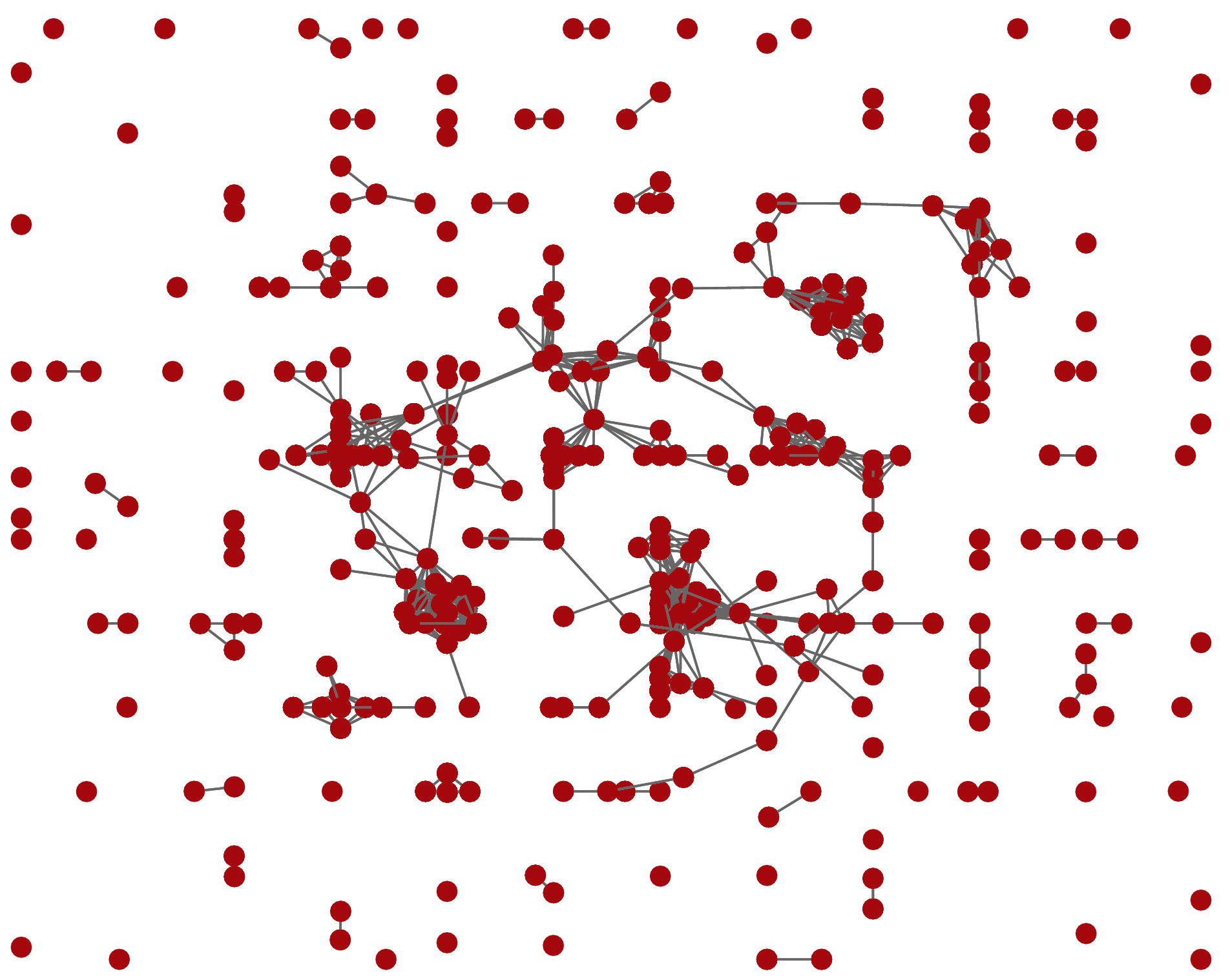} \\
    \textsf{\textbf{FR}} & \textsf{\textbf{LA}} & \textsf{\textbf{SU}} \\
    \includegraphics[height=2.4cm]{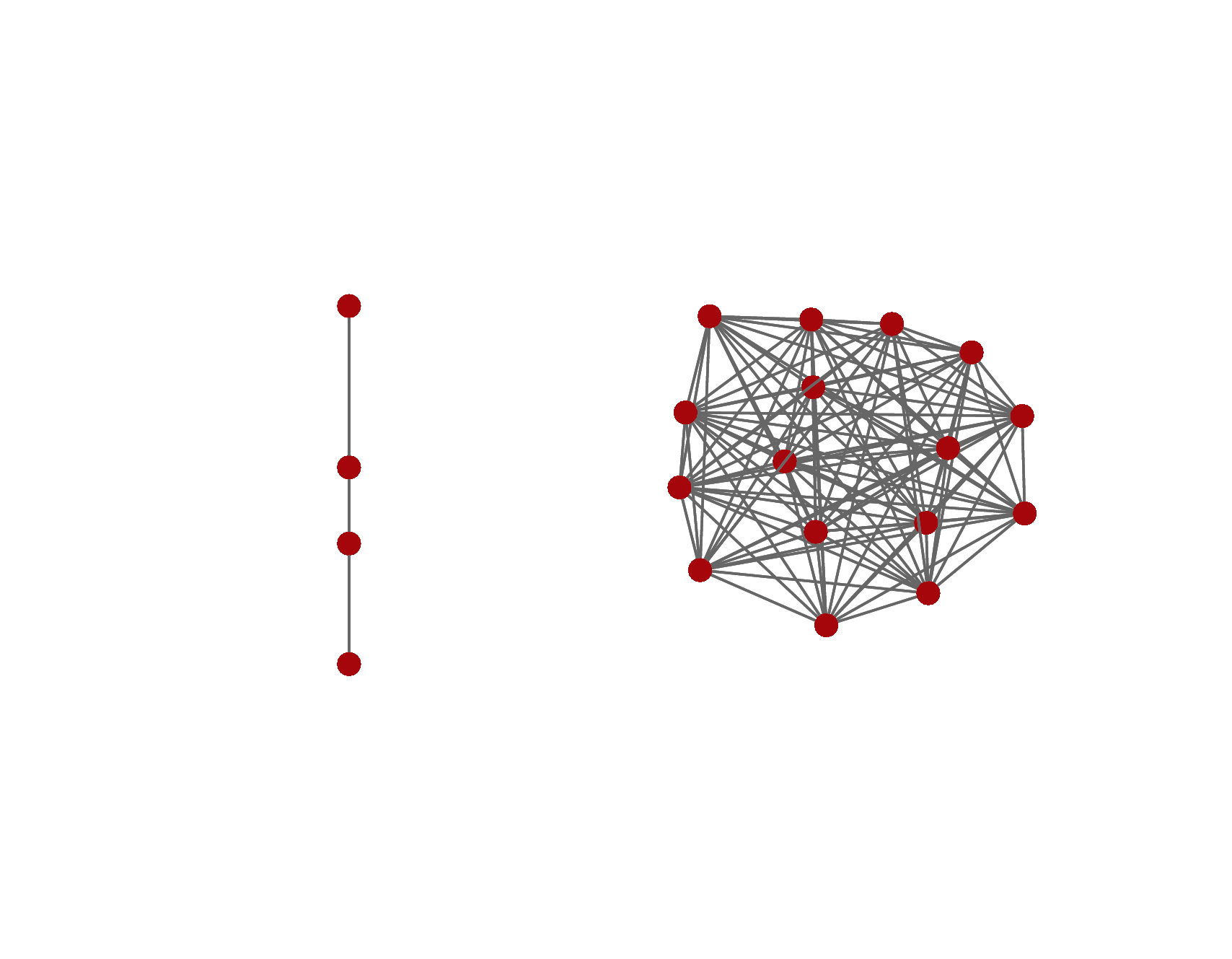} &
    \includegraphics[height=2.4cm]{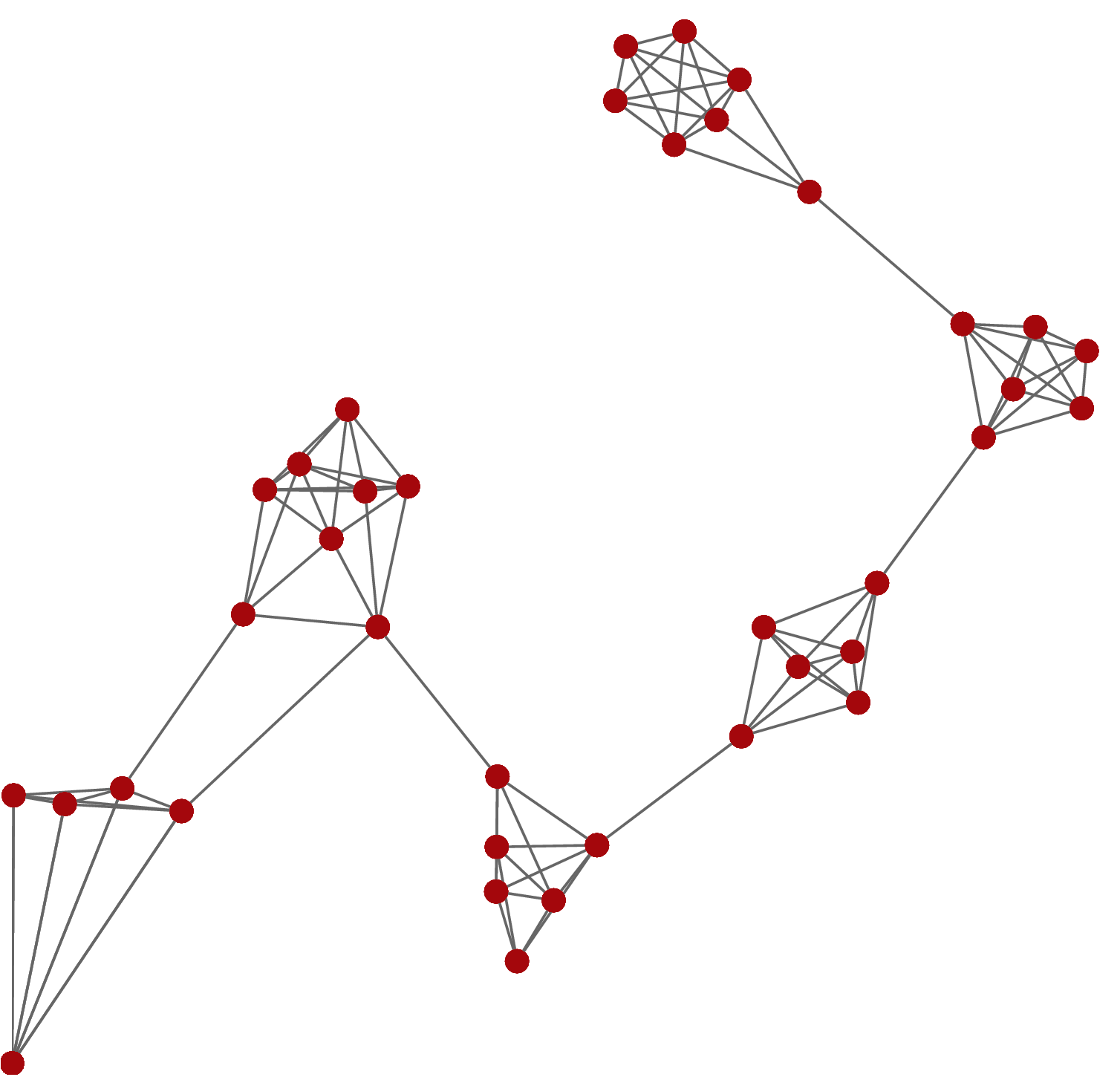} &
    \includegraphics[height=2.4cm]{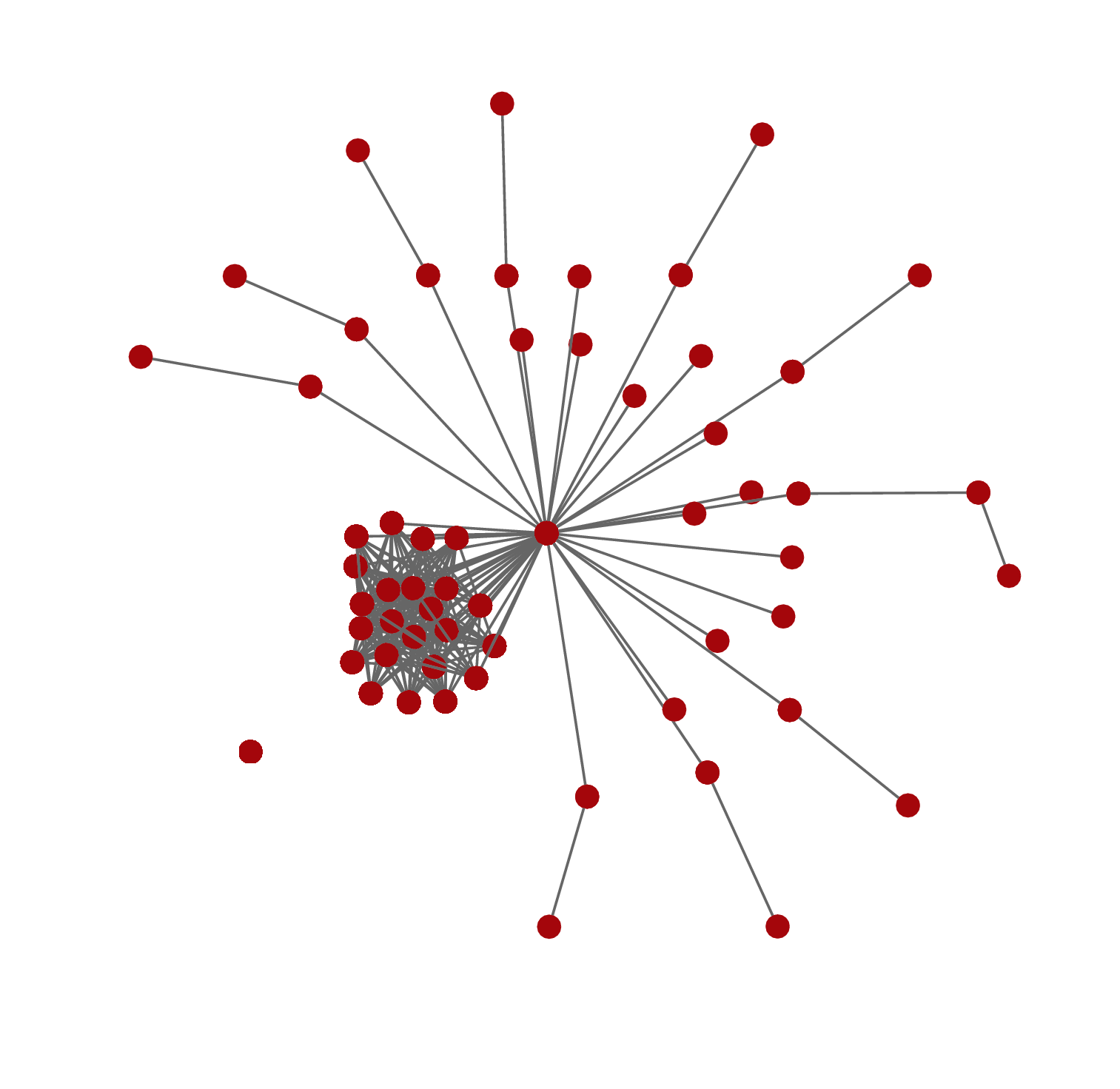} \\
    \textsf{\textbf{SN}} & \textsf{\textbf{NO}} & \textsf{\textbf{SI}} 
  \end{tabular}
  }
  \caption{ 
    The same network dataset, the Reactome network
    \citep{konect:reactome}, visualised using the six algorithms
    evaluated in this paper's experiment.  
    \textmd{
    The images are those used on the experimental website. 
    The abbreviations of the graph visualisation methods are those given in Table~\ref{tab:methods}.
    \textsf{\textbf{(FR--SN)}} Baseline
    methods. \textsf{\textbf{(NO--SI)}} SynGraphy methods. 
    }
    \label{fig:visualisations-onegraph}
  }
\end{figure}

\subsection{Experimental Setup}
The human experiment was conducted via a website\footnote{The site was available at \href{http://nohairball.west.uni-koblenz.de/}{http://nohairball.west.uni-koblenz.de/}, and is offline as of this writing.} on which practitioners
in the fields of graph mining, network analysis, and related fields were asked to
participate.  After showing the participants the definitions of the 
graph statistics
\begin{eqnarray}
  d, c, \delta, q, z, s, b \nonumber
\end{eqnarray}
as defined in Section~\ref{sec:graph-characteristics},
each participant is shown, in turn, twelve pages organised as follows.
After choosing one graph visualisation method at random (see list in
Table~\ref{tab:methods}),  
two networks are chosen at random from a set of 18 
predefined networks, and visualised using the chosen method.  Then, a 
graph statistic is chosen at random from the set of the seven mentioned graph 
statistics; the participant is asked to determine which of the two shown 
graphs has a higher or lower value for the given graph statistic.  The layout 
of the page (which graph is shown left and right), and whether the question 
is formulated as \emph{higher} or \emph{lower} is chosen randomly.  The 
user may also determine that she cannot determine the difference by 
clicking on a button for that purpose.  
The users' answers are logged, and evaluated in the following. 
The participants in the experiments were mostly from a computer
science and network science background, partly associated with the
authors' institution, and partly external. 

\begin{figure}
  \centering
  \includegraphics[width=0.6\columnwidth]{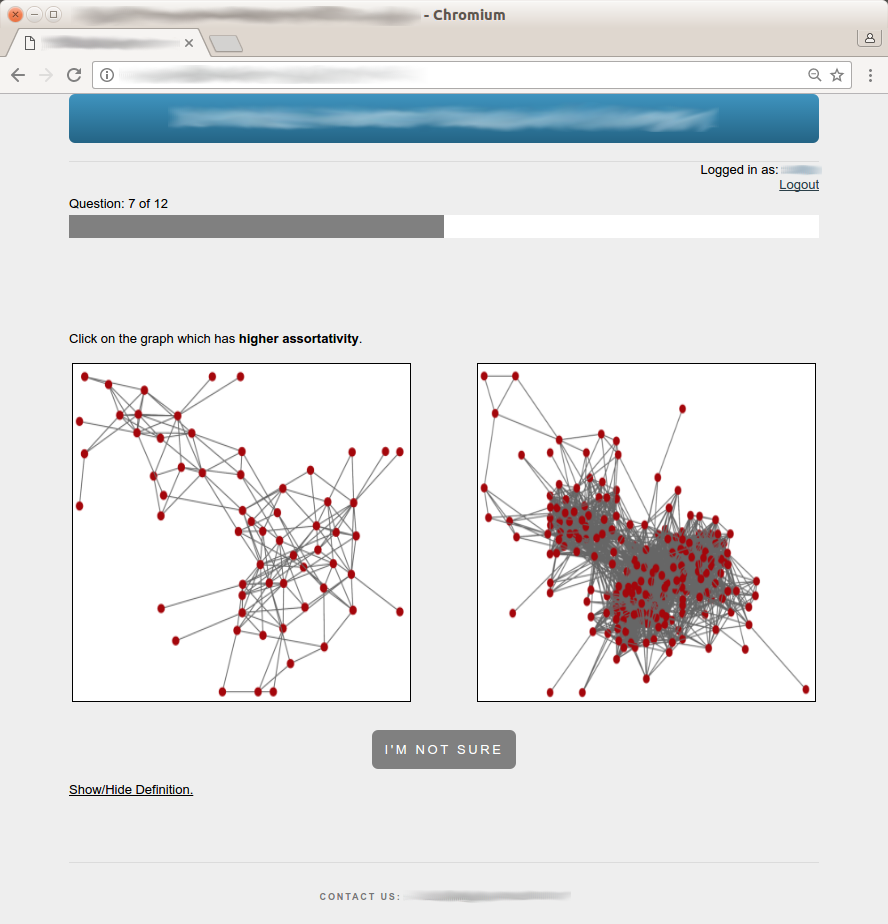}
  \caption{
    An anonymised screenshot of the human experiment.  
    \textmd{
    The screenshot
    shows a page of the experiment in which the human is tasked to
    determine visually which of the two shown networks (left or right) has
    a higher degree assortativity.  The human subject must click on the network
    to answer the question. 
    }
    \label{fig:screenshot}
  }
\end{figure}

The experiment's web page was available in mid 2016, and was
disseminated using various channels.
An anonymised screenshot of the experiment running in a browser is shown in
Figure~\ref{fig:screenshot}. For reference, 
Figure~\ref{fig:visualisations-onegraph} shows a single network
dataset visualised with all six algorithms used in the experiments. 

\begin{table}
  \caption{
    The six graph visualisation methods evaluated in the experiments. 
    \textmd{
    The last two are the variants of the SynGraphy algorithm described
    in this paper. 
    }
    \label{tab:methods}
  }
  \centering
  \begin{tabular}{ c l }
    & \textbf{Method} \\
    \hline
    \textsf{\textbf{FR}} & Fruchterman--Reingold (\citeyear{b870}) \\
    \rowcolor{tabbgd}\textsf{\textbf{LA}} & Laplacian embedding \citep[see e.g.][]{b405} \\
    \textsf{\textbf{SU}} & Uniform vertex sampling\textsuperscript{1} \\
    \rowcolor{tabbgd}\textsf{\textbf{SN}} & Node sampling\textsuperscript{1} \\
    \textsf{\textbf{NO}} & SynGraphy with empirical statistic distribution\textsuperscript{1} \\
    \rowcolor{tabbgd}\textsf{\textbf{SI}} & SynGraphy with size-independent statistics\textsuperscript{1} \\
    \hline
    \multicolumn{2}{p{8cm}}{
      \textsuperscript{1}These methods are followed by the use of the
      Fruchterman--Reingold method for drawing the resulting small
      graph. 
    }
  \end{tabular}
\end{table}

\begin{figure}
  \centering
  \includegraphics[width=0.6\textwidth]{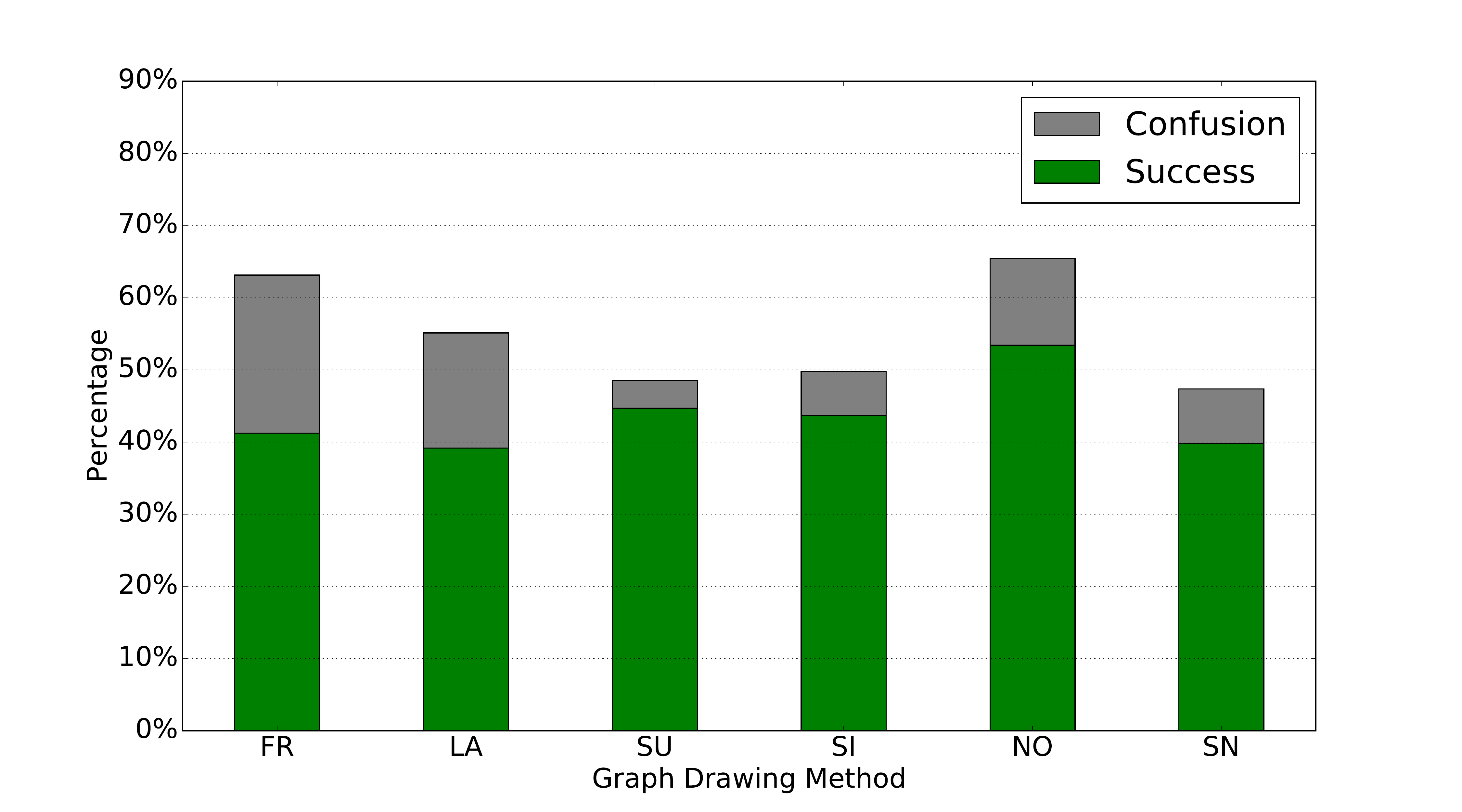}
  \caption{
    The experimental results showing the success rate and confusion
    rate of the six evaluated graph visualisation methods.  
    \textmd{
    The
    rightmost two, \textsf{\textbf{NO}} and \textsf{\textbf{SI}}, are the variants of the
    SynGraphy algorithm described in this paper. 
    }
    \label{fig:plot-method}
  }
\end{figure}

\begin{figure}
  \centering
  \includegraphics[width=0.6\textwidth]{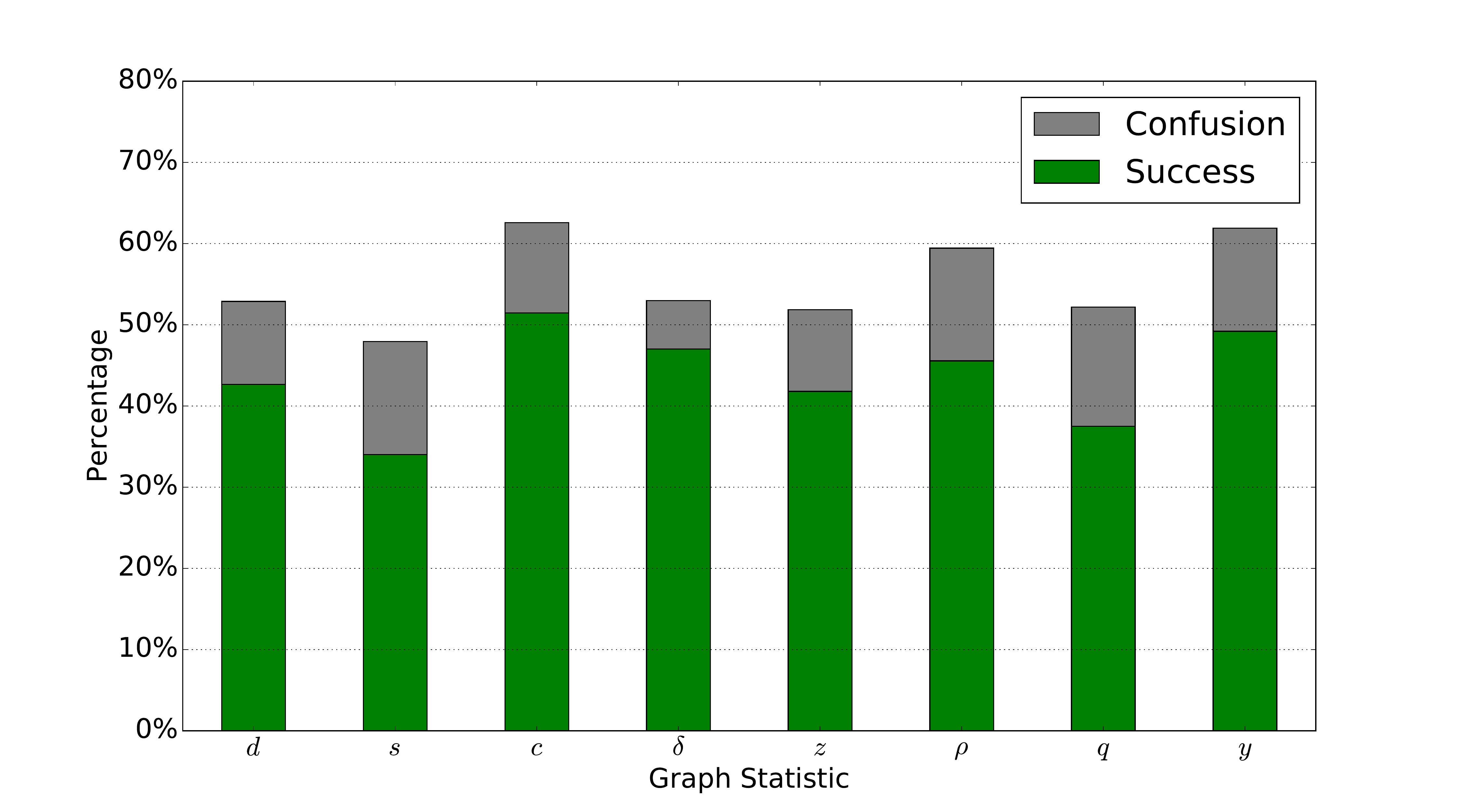}
  \caption{
    The experimental results showing the success rate and confusion rate
    for each of the network statistics employed in the experiments. 
    \textmd{
    In particular, the plots shows that as measured over all graph
    visualisation methods, certain network statistics are very difficult
    to discern. 
    }
    \label{fig:plot-stat}
  }
\end{figure}

\begin{figure}
  \centering
  \includegraphics[width=0.6\textwidth]{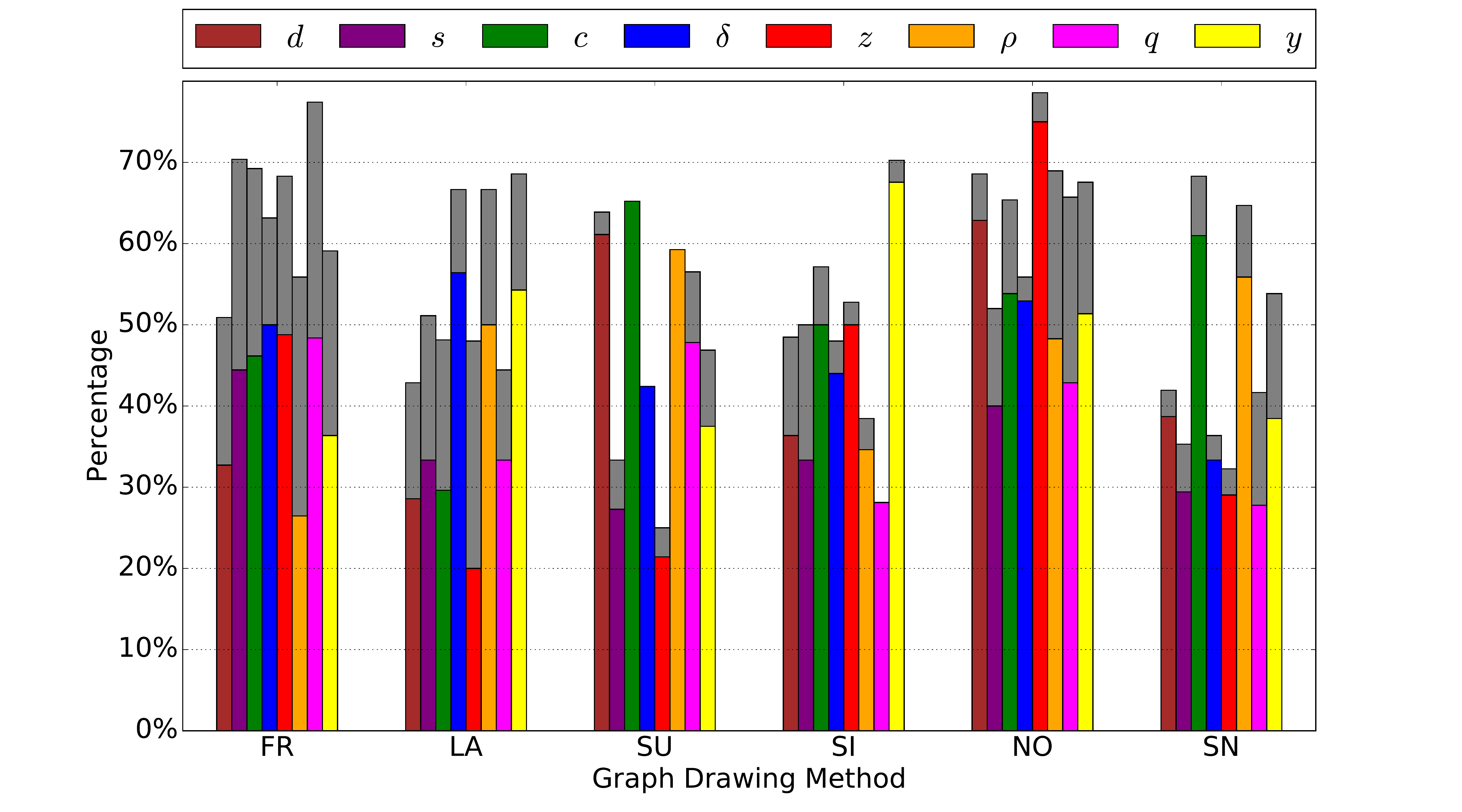}
  \caption{
    The experimental results broken down both by graph visualisation
    method and network statistic.
    \textmd{
    The grey bars represent the confusion rate. 
    }
    \label{fig:plot-stat-method}
  }
\end{figure}

The experiment was divided into two parts. In the first part, we 
show to the experimental subject the definitions of the various network
properties and show two example graphs with high
and low values of each properties. 
In the second part, some random graphs (generated using the methods
discussed above) were shown to the users. The user had to identify the
graph with a low or high value of a certain property. 
The six graph visualisation methods used in the experiment are given in
Table~\ref{tab:methods}. 
For the uniform vertex sampling method, we chose a random subset of
the nodes, and keep all edges connected to at least one of them
\citep{clausetNotes,hu:graph-sampling}. 
For node sampling, we chose a random subset of all nodes, and
keep all edges connecting two of them. 

\subsection{Results}
From the logs of the experiments, we determine for each page view
whether the person was correct or incorrect in their assessment of the
given network statistic for the two given graphs visualised with the
given method. 
In total, 3,049 questions where answered by 79 unique identified
participants, plus an unknown number of anonymous
participants.\footnote{Participants were given the possibility to
  optionally enter their name into the system.} 
Each question is thus answered in exactly one of three ways:  correct,
incorrect, or skipped.  Thus, we collect three
counts:
\begin{itemize}
\item $N^+$, the number of correct answers,
\item $N^-$, the number of incorrect answers,
\item $N^0$, the number of skipped answers. 
\end{itemize}
The total number of questions shown is then $N^+ + N^- + N^0$. 
From these numbers, we compute the success rate $R^s$ and the confusion
rate $R^c$ as follows: 
\begin{eqnarray}
  R^s = N^+ / (N^+ + N^- + N^0), \qquad R^c = N^0 / (N^+ + N^- + N^0)
\end{eqnarray}
The experimental results are shown in figures
\ref{fig:plot-method}, \ref{fig:plot-stat} and \ref{fig:plot-stat-method}.  

\vspace{0.1cm}
\noindent
\textbf{Success Rate by Visualisation Method.}
Figure \ref{fig:plot-method} shows the overall success rate
for each graph visualisation method used in the experiment, including
the two SynGraphy variants. 
The result
shows that the overall success rate of the SynGraphy variant based on
normalisation (\textsf{\textbf{NO}}) is significantly higher than all 
the other methods used in the experiment ($\chi^2$ test,
$p=0.000287$). The SynGraphy variant based on statistics that are by
construction independent of size (\textsf{\textbf{SI}})
does not show statistically better results than the node sampling
(\textsf{\textbf{SN}}) method ($\chi^2$ test, $p=0.447$) and the uniform vertex sampling
(\textsf{\textbf{SU}}) method ($\chi^2$ test, $p=0.984$).  
The fact that the empirical normalisation method (\textsf{\textbf{NO}})
performs much better than the one using predetermined normalised
statistics (\textsf{\textbf{SI}}) is also enlightening, as it suggests
that statistics such as the clustering coefficient, which are
size-independent in theory by construction, are not necessarily so in
practice, and that in practice it is more effective to normalise
subgraph count statistics using available empirical data from actual
network datasets. 

\vspace{0.1cm}
\noindent
\textbf{Success Rate by Network Statistic.}
Comparing the individual network statistics with each other, it appears
that the task of identifying structural network properties is, as a
general rule, a difficult one.  As seen in Figure~\ref{fig:plot-stat}, only a single
statistic reaches an overall success rate (over all visualisation
methods combined) of more than 50\%, this being the clustering
coefficient.  This may be explained by the better knowledge of the
clustering coefficient of many participants, who, despite the
explanations given at the start of the experiments, may not be familiar
with less well-known network statistics such as the degree assortativity, and
may thus have a harder time recognising it in graph drawings. 
It can also be observed that non-normalised statistics such as the raw
number of wedges ($s$) have low success rates -- justifying the need for
normalised statistics such as the clustering coefficient. 

\vspace{0.1cm}
\noindent
\textbf{Confusion Rate.}
In Figure \ref{fig:plot-method}, we may also compare the confusion rate of
the different graph visualisation methods.  The interpretation of the
confusion rate is ambivalent:  A high confusion rate may indicate that a
particular graph visualisation method is confusing for the user, but at
the same time a graph visualisation method should avoid misleading users
-- it is better if the user is undecided, than for her to be wrong.
Thus, the confusion rate should not be taken as an indicator of the quality
of a graph visualisation method, but rather as a measure of how
convincing the resulting plots look.  Thus, a graph visualisation method
with a low confusion rate may lead to the user often thinking that they
are able to recognise the properties of the graph, regardless of whether
they are right or wrong.  That being said, the smallest confusion rates
are achieved by the two sampling methods (\textsf{\textbf{SU}}, \textsf{\textbf{SN}}), and the
highest by the direct drawing methods (\textsf{\textbf{FR}}, \textsf{\textbf{LA}}).  The
two SynGraphy variants then achieve an intermediate confusion rate. 
To be precise, 
the size-independent SynGraphy (\textsf{\textbf{SI}}) method has a significantly lower
confusion rate than the Fruchterman--Reingold (\textsf{\textbf{FR}}) method 
($\chi^2$ test, $p=2.78{\times}10^{-7}$), and also than the four baseline methods combined
($\chi^2$ test, $p=0.00356$). 

\vspace{0.1cm}
\noindent
\textbf{Individual Results.}
The results shown in Figure \ref{fig:plot-stat-method} allow us to make
more detailed remarks about the performance of the individual graph
visualisation methods with respect to individual network statistics. 
We observe that the SynGraphy variant based on empirically derived
normalisation (\textsf{\textbf{NO}}) gives the 
highest success rate for the number of claws and density, while for
other network statistics, other visualisation methods than SynGraphy
perform better, although no single one is better for many network
statistics.  In particular, the plain Fruchterman--Reingold method has
the highest success rate for the number of wedges and squares, the
Laplacian embedding has the highest success rate for the diameter,
uniform vertex sampling has the highest success rate for the clustering
coefficient and assortativity, and finally node sampling has the highest
success rate for the bipartivity coefficient.  Thus, it seems that if a
graph visualisation is needed in which a particular network property
should be emphasized, then we cannot claim that the SynGraphy method is
the best choice in all cases.  It should also be noted that if a
particular property such as bipartivity is to be emphasized, then other,
more specific methods may be chosen. 

\subsection{Runtime Experiments}
\begin{figure}
  \centering
  \includegraphics[width=0.60\textwidth]{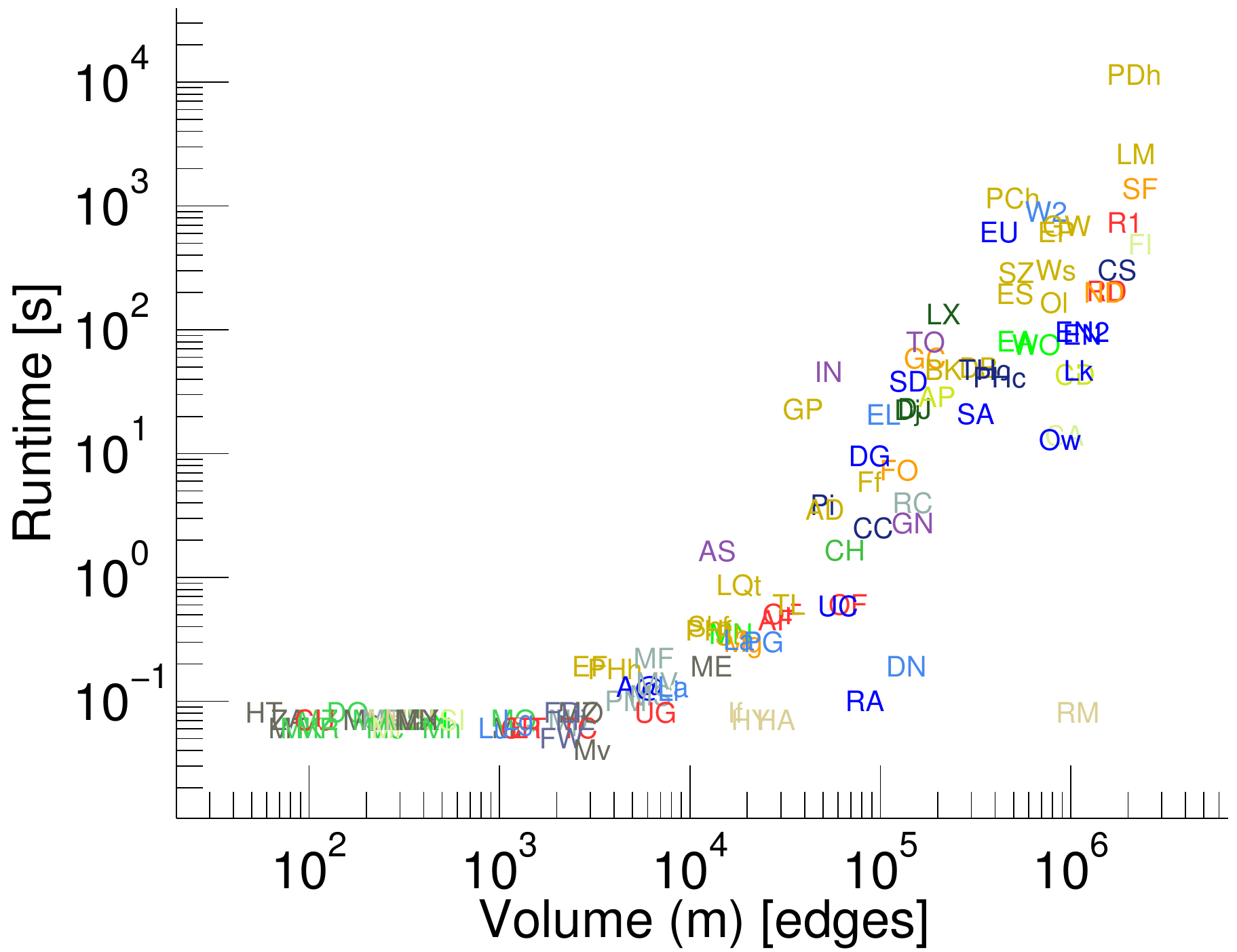}
  \caption{
    The runtime of the SynGraphy algorithm for each tested network
    dataset, as a function of the number of edges in the network. 
    \textmd{
    For small networks (number of edges $m < 10^4$) the runtime is
    dominated by the constant runtime of the scaling down, small graph
    generation and graph drawing steps of the SynGraphy algorithm, while for larger networks ($m > 10^4$), the
    runtime is empirically quadratic due to the subgraph count
    computation step, in accordance with the theoretical
    model described in the text. 
    }
    \label{fig:runtime}
  }
\end{figure}

In order to assess the runtime of the method empirically, we show the
runtime of the SynGraphy algorithm in Figure~\ref{fig:runtime}.  The two
variants of SynGraphy have, for all practical purposes, the same
runtime, as they differ only in the step of scaling down the network
statistics, whose runtime is constant and very small, i.e., it does not depend on the
network size.  The amount of memory taken up by a sparse graph is
proportional to the number of edges rather than the number of nodes, and
thus we measure the runtime in function of the number of edges, i.e.,
the graph's volume $m$. 
We can derive the runtime of the SynGraphy algorithm in function of the
input graph's volume $m$ as follows. 
The different steps of the algorithm are in turn:  (i)~computation of
subgraph counts, (ii)~scaling down of subgraph counts, (iii)~generation of
the synthetic graph, and (iv)~generation of the graph layout using the
Fruchterman--Reingold algorithm.  Steps (ii--iv) have runtime independent
of the graph's number of edges $m$, as they are performed on data of
fixed size, and thus have a formal runtime of $O(1)$.  Step~(i), the
counting of subgraphs, is the only part of the algorithm depending on $m$,
and its actual runtime is dominated by the computation of the counting
of squares (4-cycle subgraphs), which is quadratic.  Thus, the overall
runtime of SynGraphy is expected on theoretical grounds to be $O(m^2)$.
The empirical data as shown in Figure \ref{fig:runtime} is consistent
with that runtime of $O(m^2)$. 

\subsection{Varying $n'$}
In order to assess the influence of the size $n'$ of the generated small
synthetic graphs on the quality of the summarisation, we generate graph
summarisations for varying $n'$ using a single network, the email
network of Enron \citep{konect:klimt04} already used in
Figure~\ref{fig:hairball-gallery}, visualised using the SynGraphy
variant based on 
the empirically learned normal distribution of statistics (\textsf{\textbf{NO}}).
The
resulting graphs are shown in Figure~\ref{fig:scalability}, for varying $n'$
between a value of $n'=10$ up to a value of $n'=1000$. 
\newcommand{\scalabilityHeight}{0.80cm}
\begin{figure}
  \makebox[\textwidth]{
  \begin{tabular}{ c c c c c c c }
    \includegraphics[height=\scalabilityHeight]{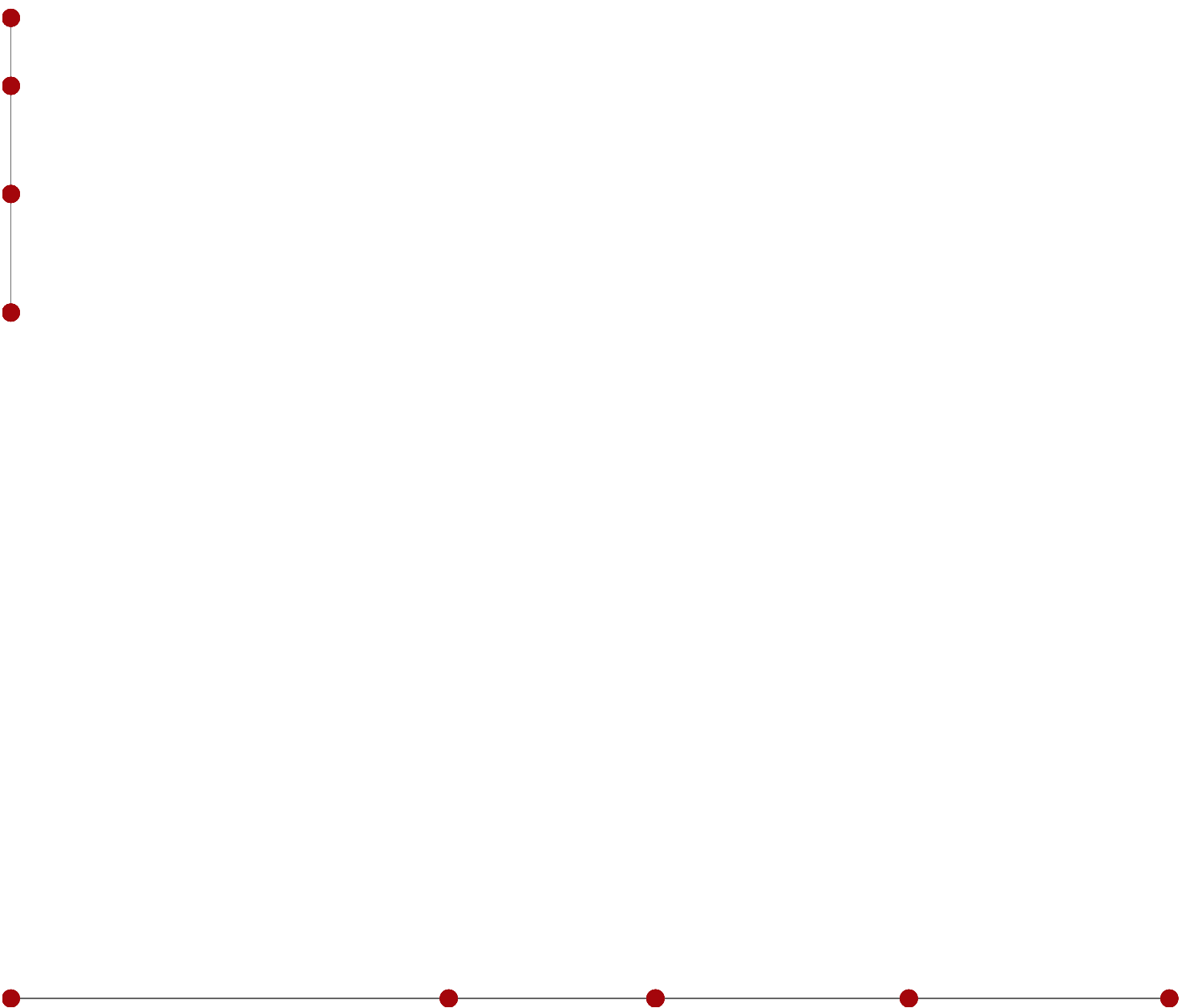} &
    \includegraphics[height=\scalabilityHeight]{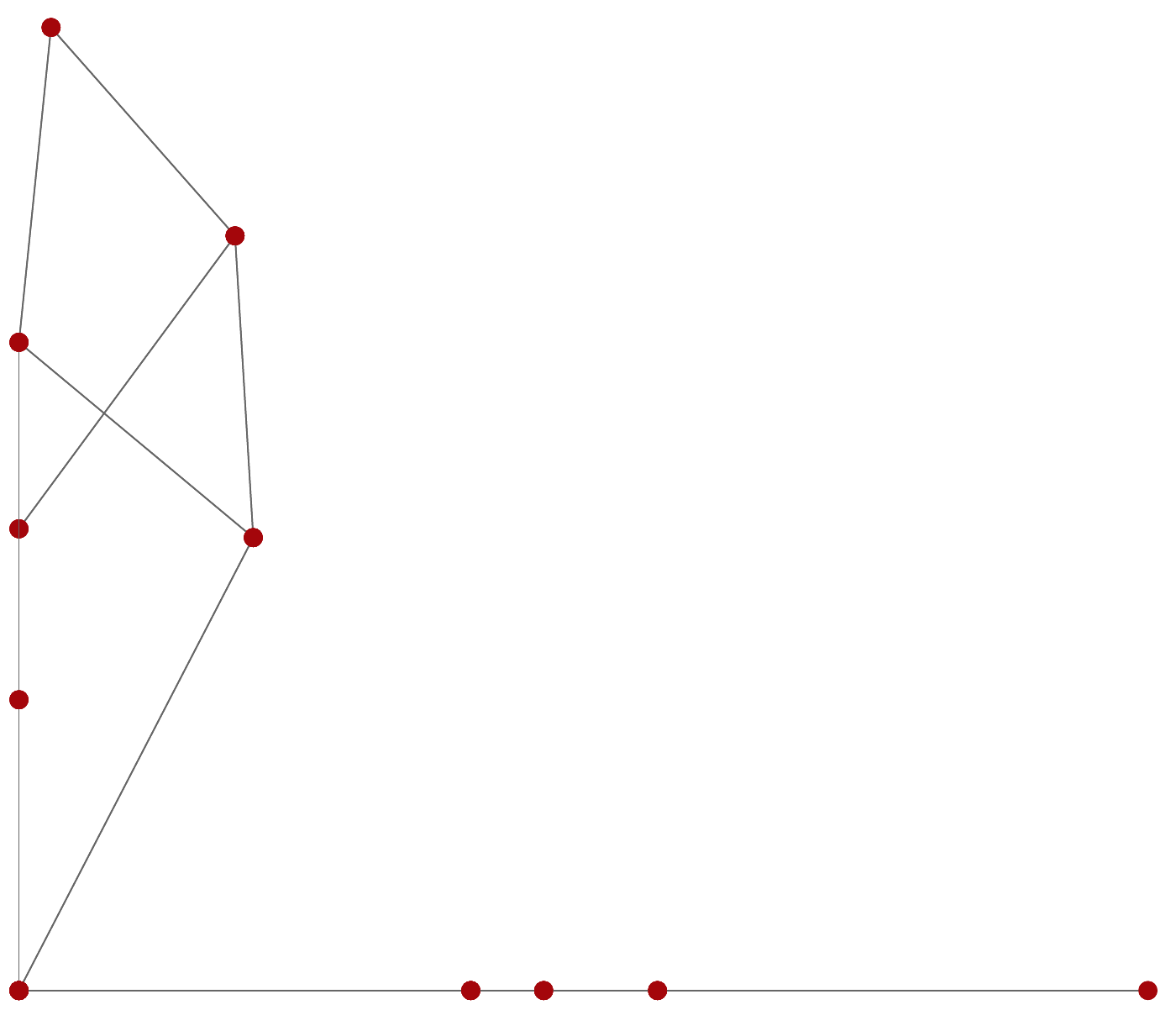} &
    \includegraphics[height=\scalabilityHeight]{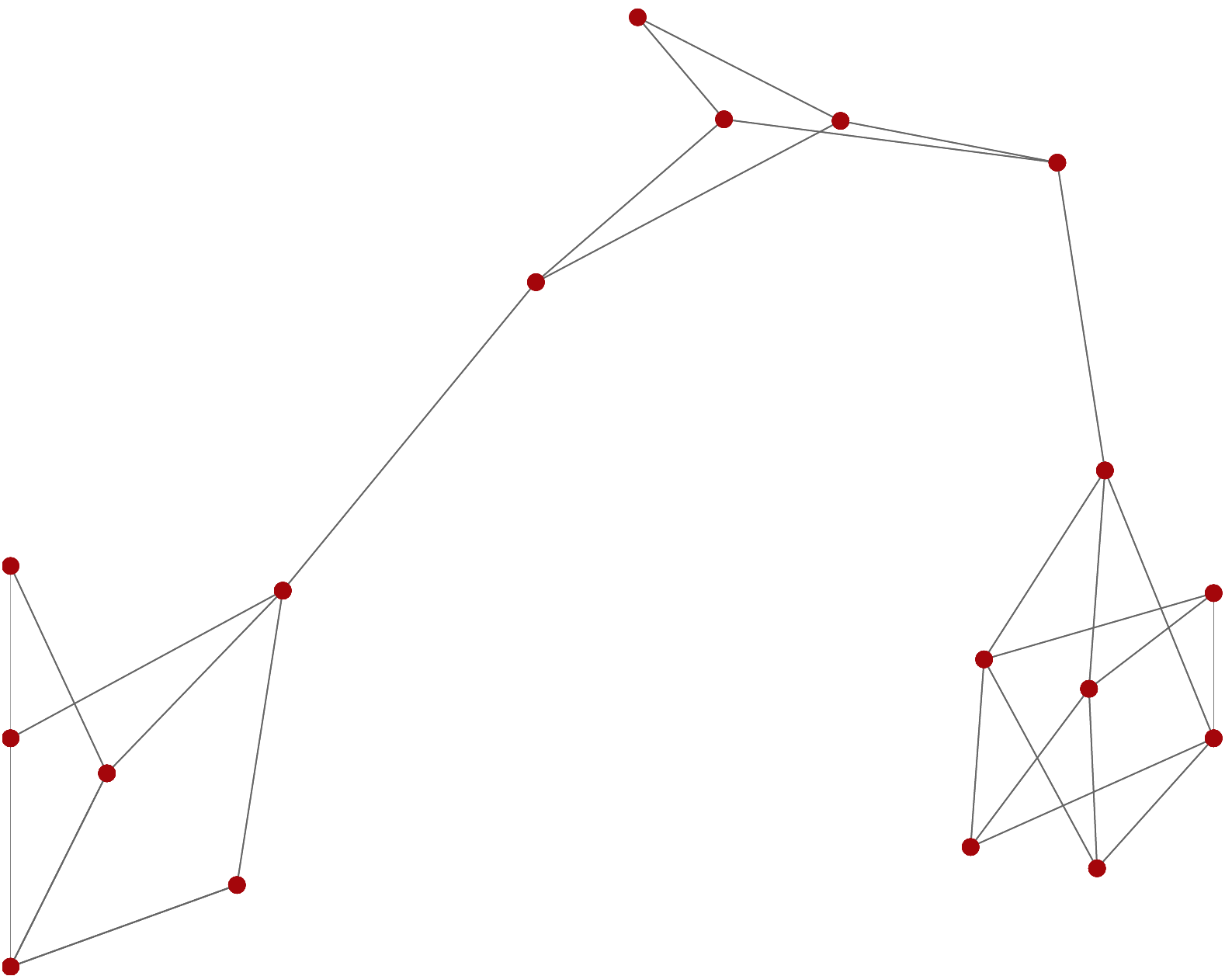} &
    \includegraphics[height=\scalabilityHeight]{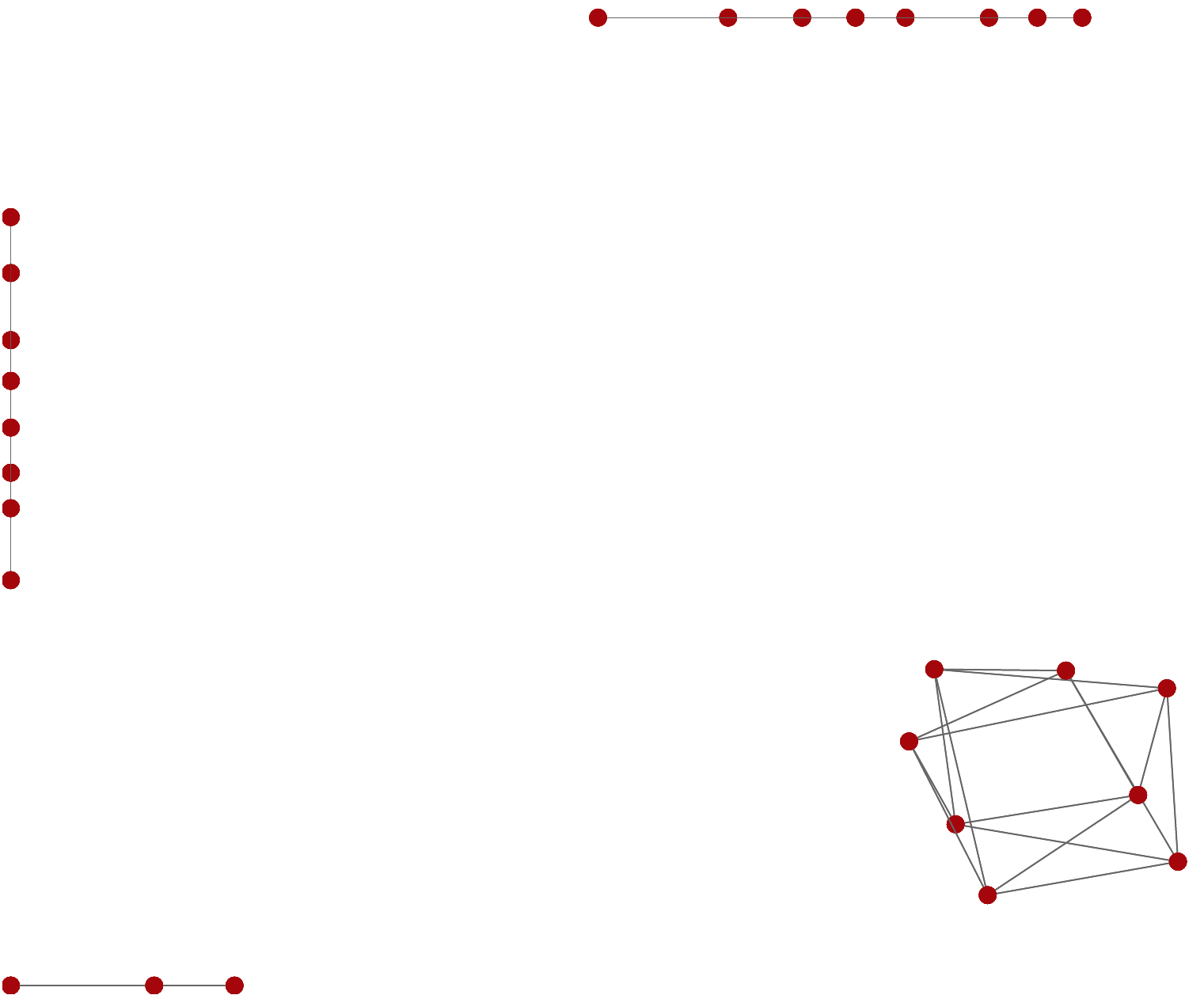} &
    \includegraphics[height=\scalabilityHeight]{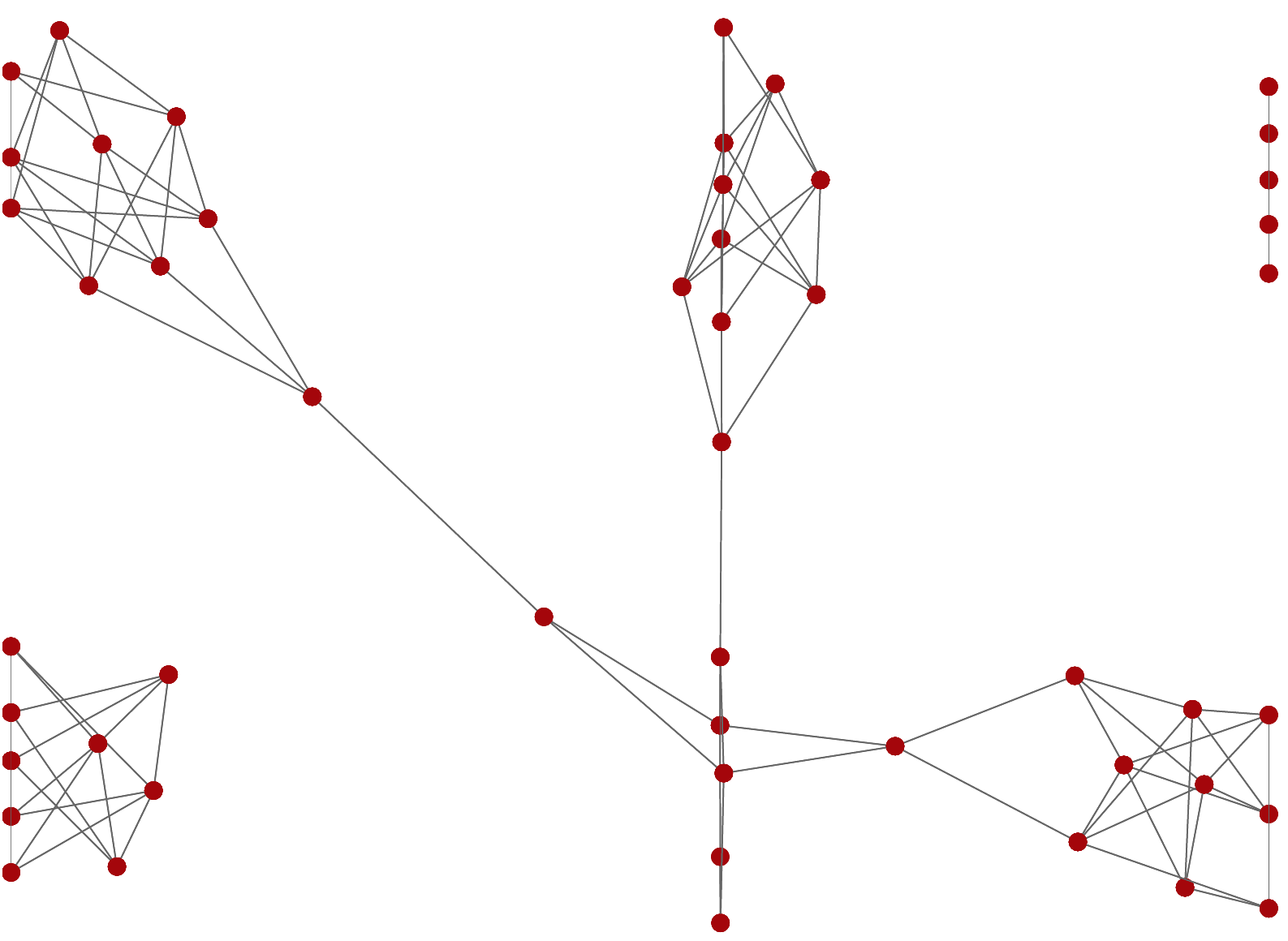} &
    \includegraphics[height=\scalabilityHeight]{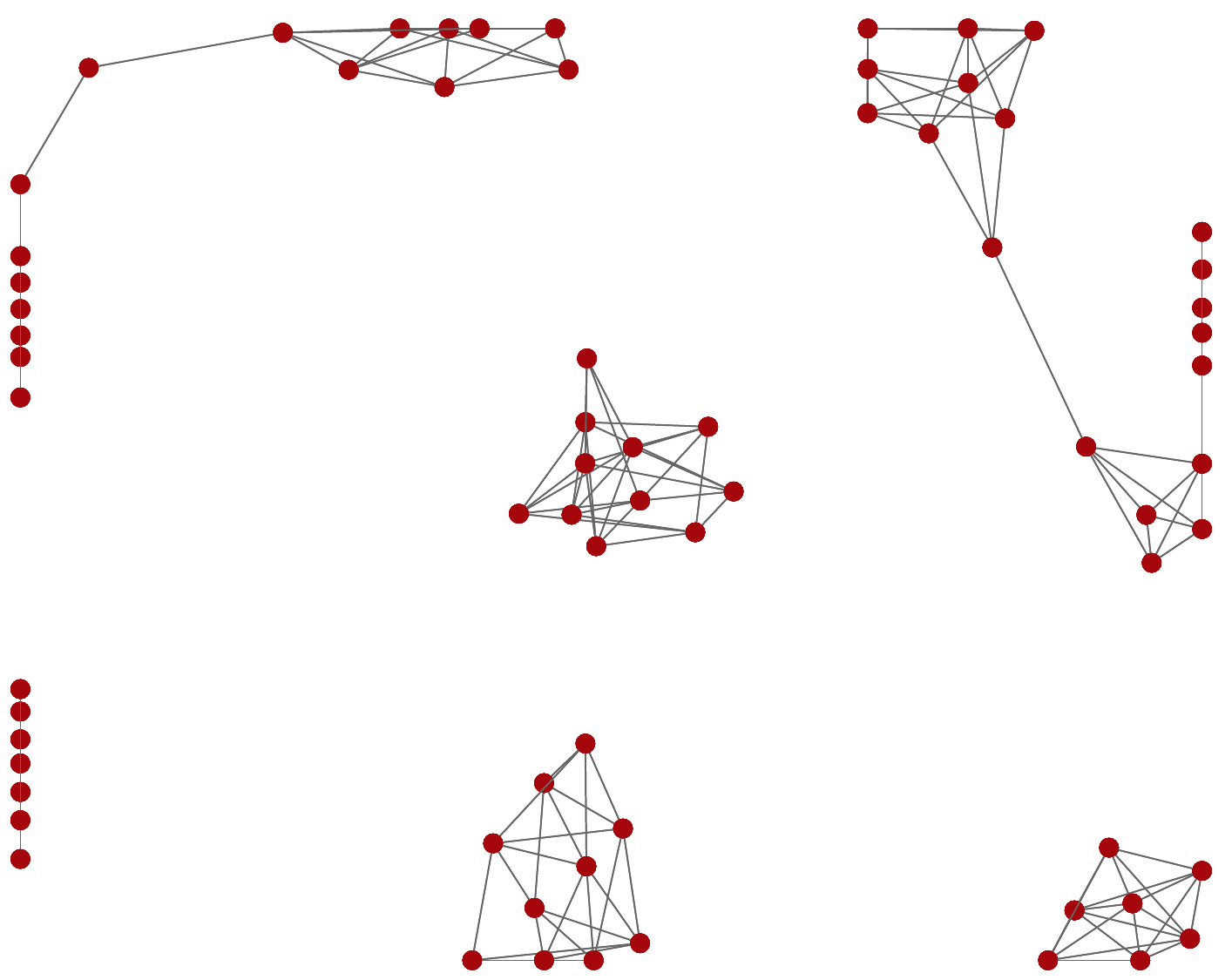} &
    \includegraphics[height=\scalabilityHeight]{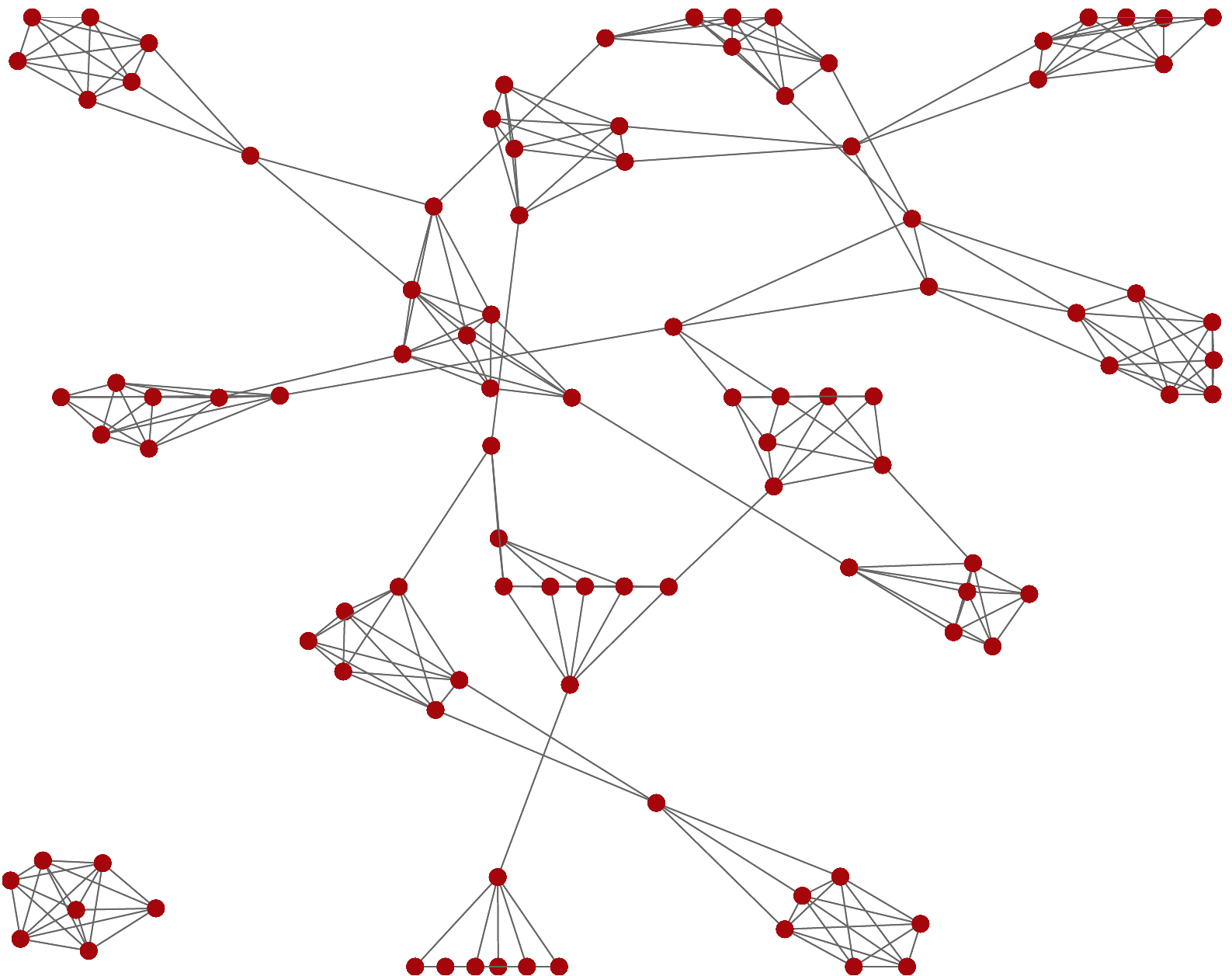} \\
    $n'=10$ & $15$ & $20$ & $30$ & $50$ & $70$ & $100$ \\
    \includegraphics[height=\scalabilityHeight]{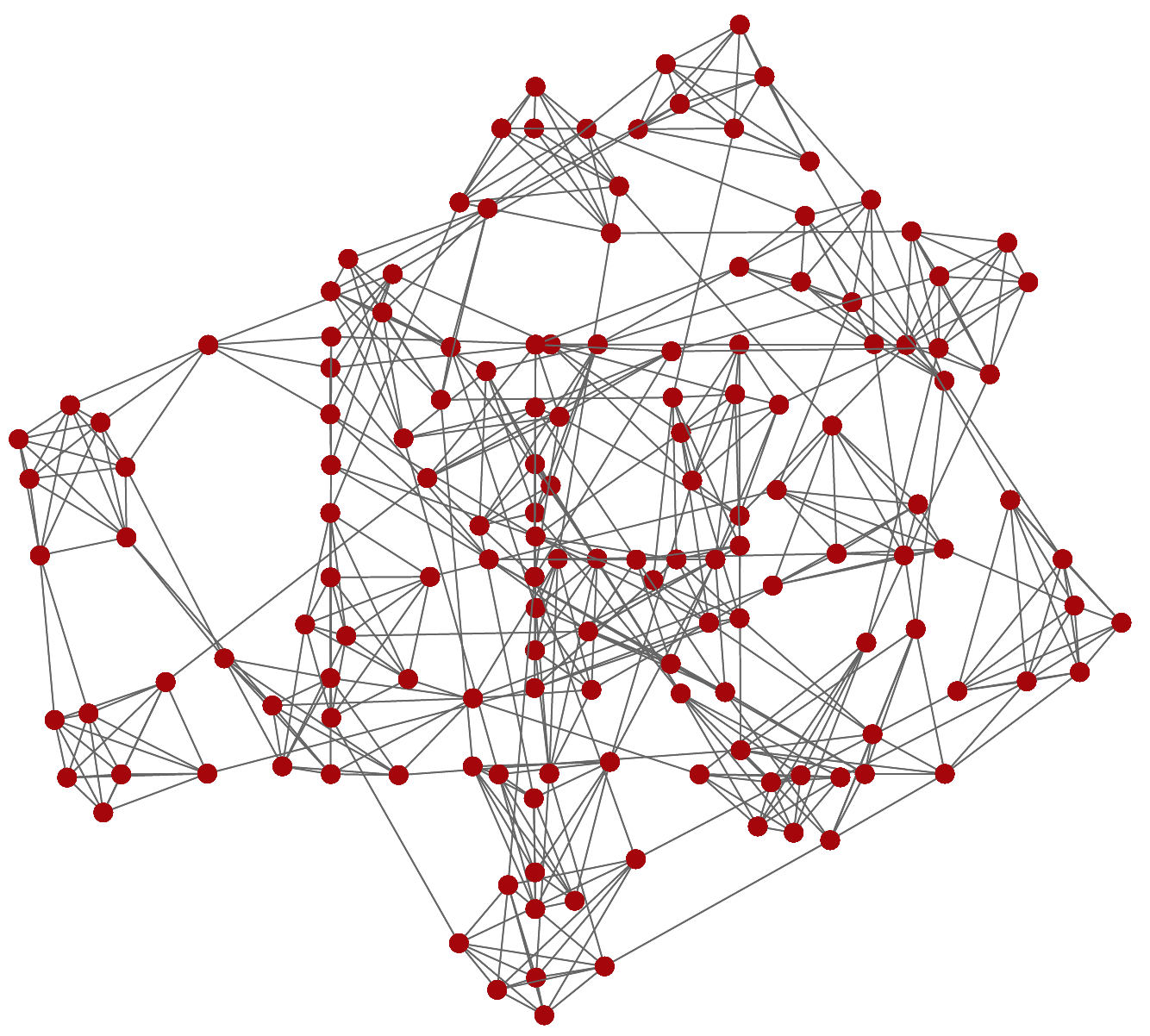} &
    \includegraphics[height=\scalabilityHeight]{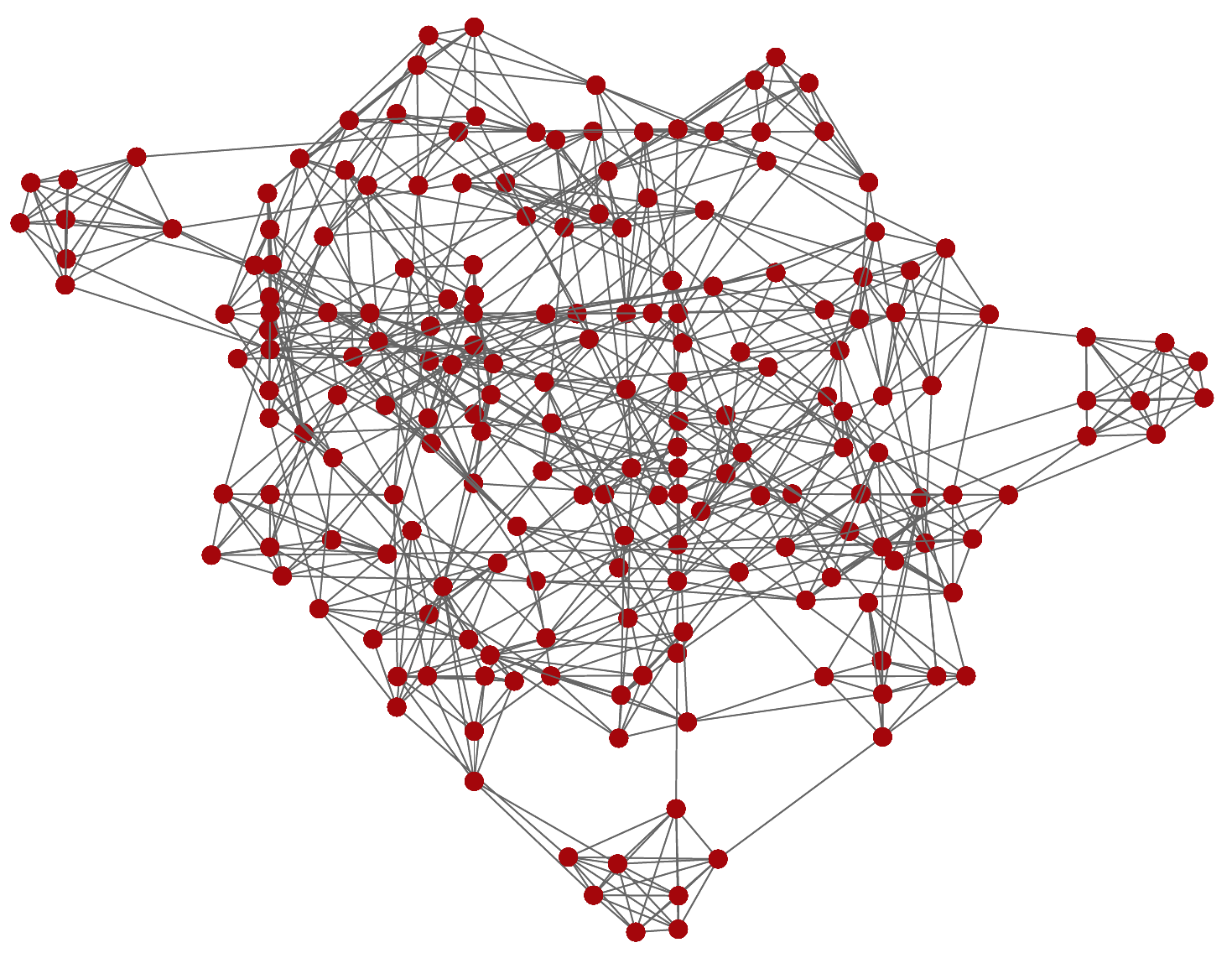} &
    \includegraphics[height=\scalabilityHeight]{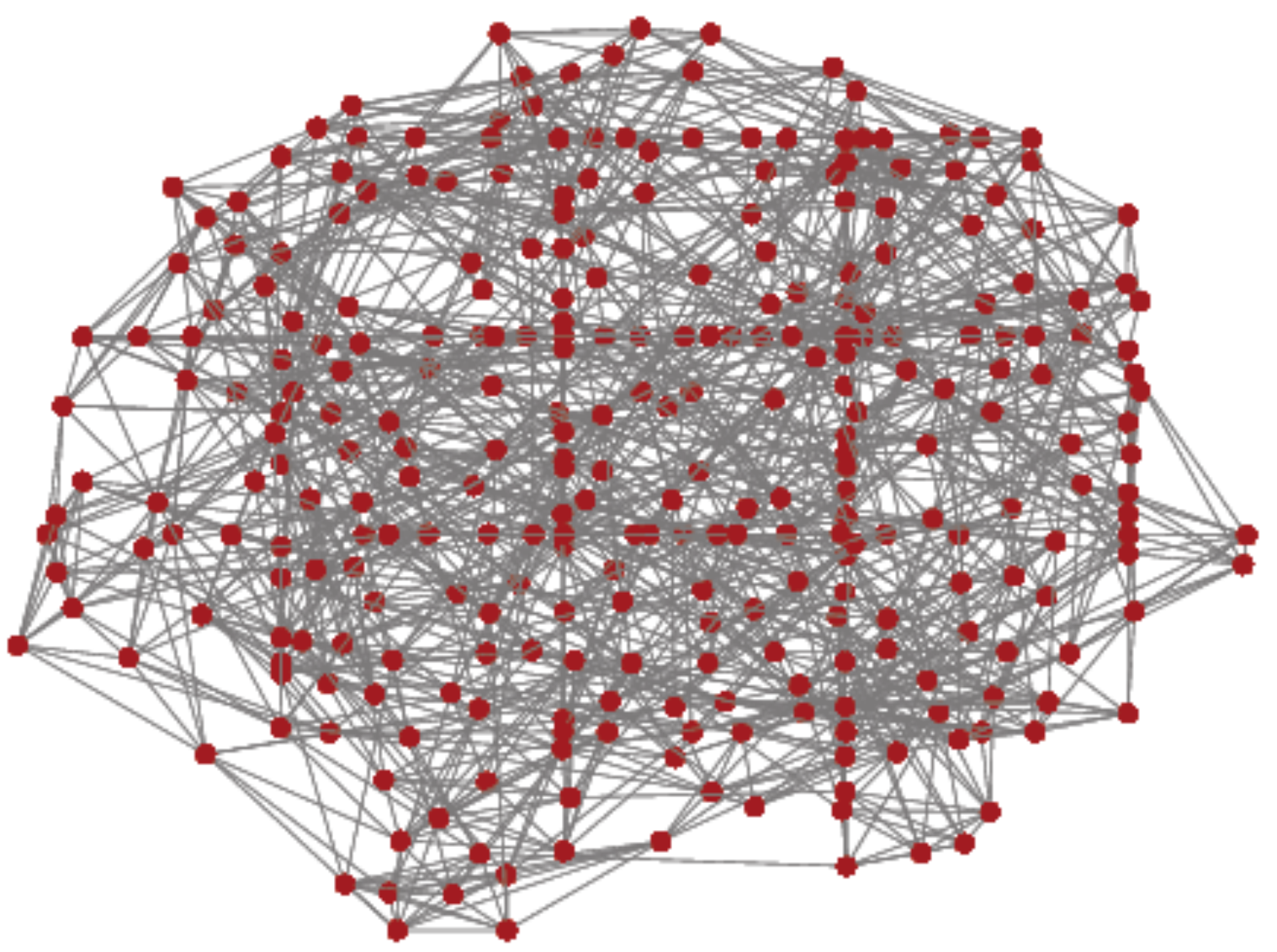} &
    \includegraphics[height=\scalabilityHeight]{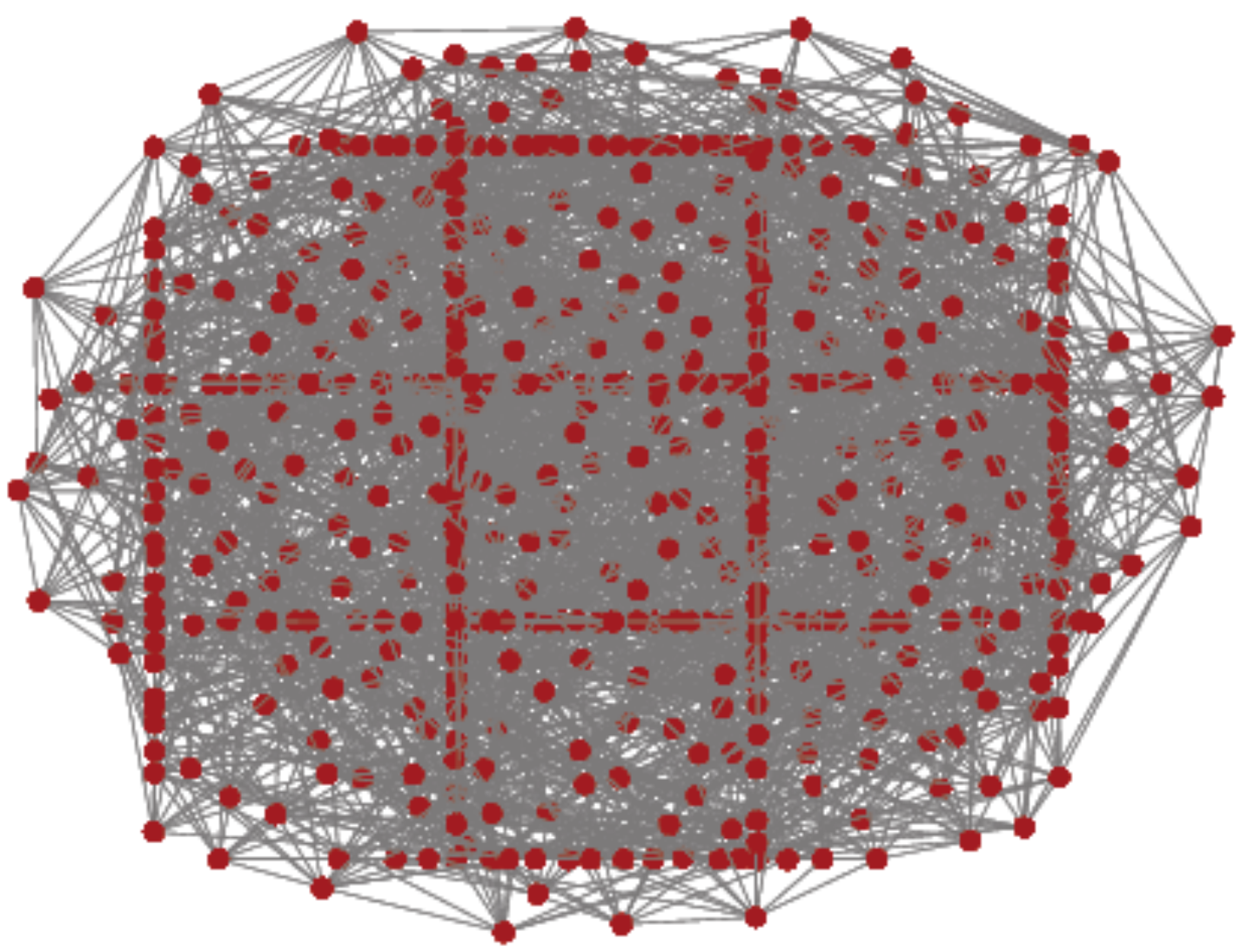} &
    \includegraphics[height=\scalabilityHeight]{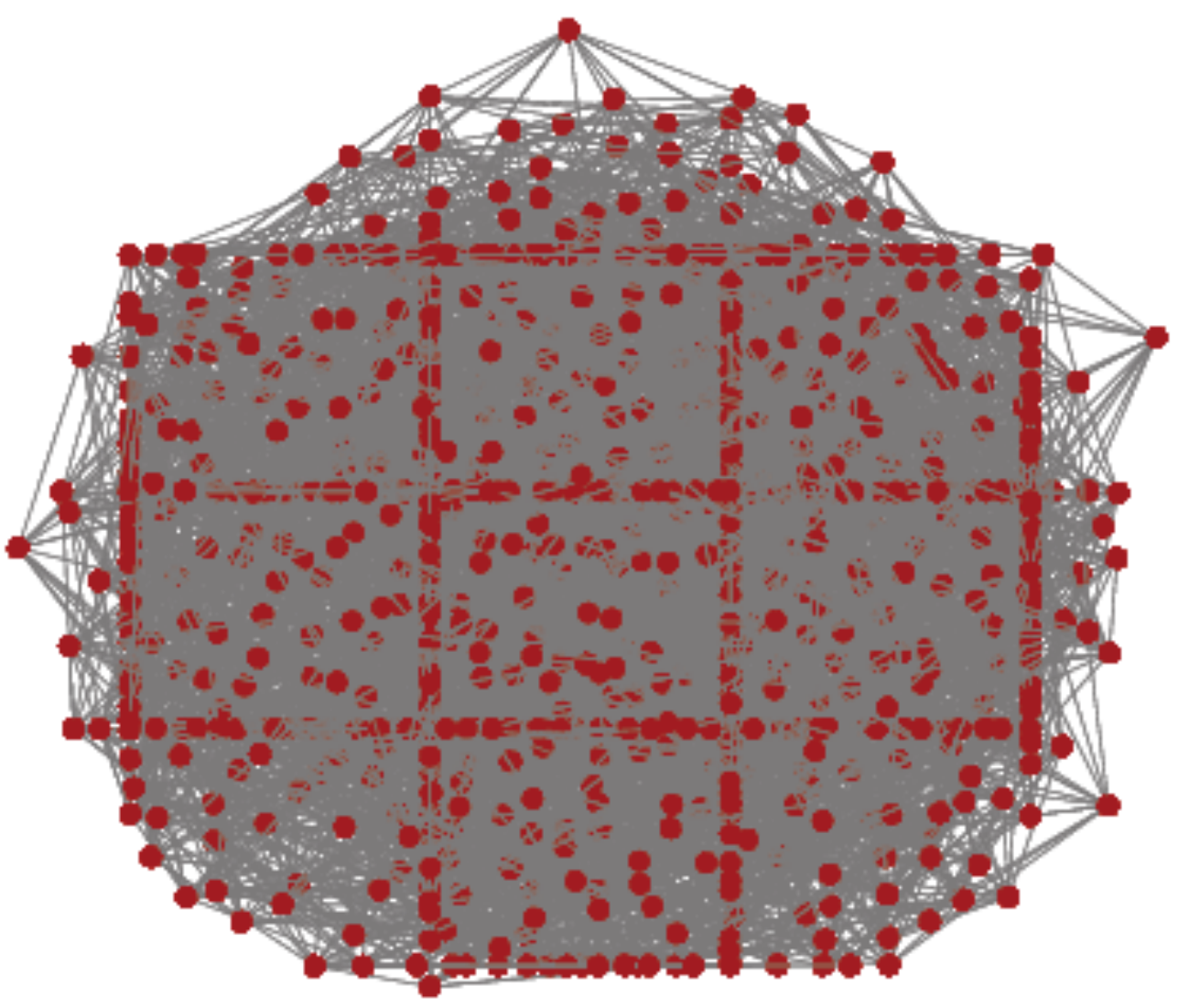} &
    \includegraphics[height=\scalabilityHeight]{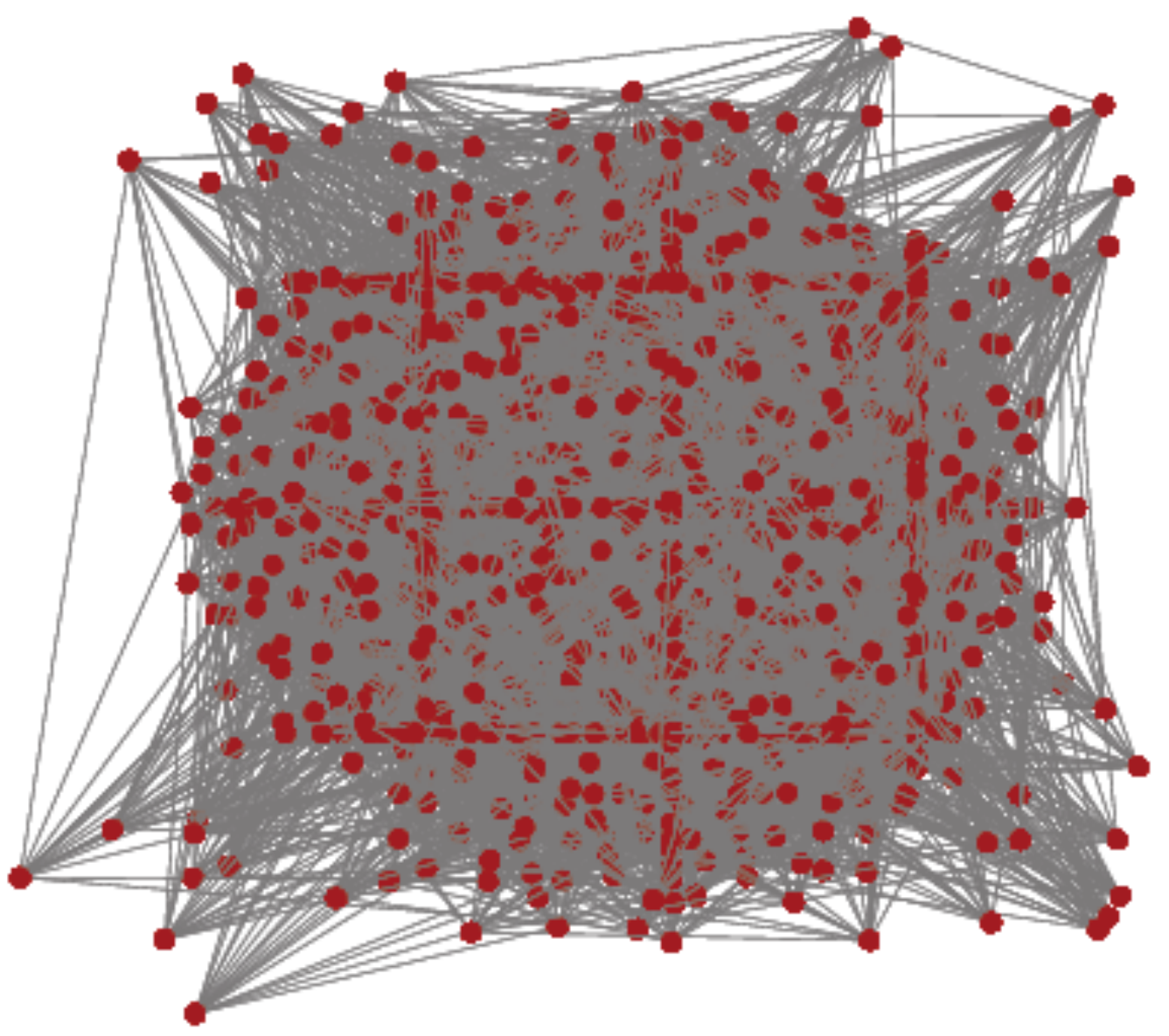} \\
    $150$ & $200$ & $300$ & $500$ & $700$ & $1000$ 
  \end{tabular}
  }
  \caption{
    The email dataset from \citep{konect:klimt04} summarised with a
    varying number of nodes $n'$ in the resulting graph, using SynGraphy
    with the empirical statistic distribution (\textsf{\textbf{NO}}).
    \textmd{
    When $n'$ is  
    very small, the generated graphs are uninformative, while for too
    large $n'$, the drawings converge to a ``hairball''-like drawing.
    For intermediate values of $n'$ in the range $70 \leq n' \leq 100$,
    the drawings strike a balance between enough detail and enough space
    to make the network structure evident. 
    }
    \label{fig:scalability}
  }
\end{figure}
We observe that for very small $n'$, the small number of drawn nodes and
edges is not enough to convey any meaningful information about the
network, and thus the summarisation of the original graph is not a
useful representation of the original network.  For too large values of
$n'$, the drawn graphs approach a ``hairball''-like image, and thus they
too are not useful representations of the graph to be summarised.
Intermediate values in the approximate range $70 \leq n' \leq 100$ give
an acceptable trade-off between the two conditions, leading us to
recommend a value of 80 for $n'$.  This is also the value used in the
previous experiments, as explained in the previous section. 

\section{Discussion}
\label{sec:conclusion}
We have seen that in the overall evaluation the proposed SynGraphy
variant based on the empirical distribution of graph statistics
(\textsf{\textbf{NO}}) has the highest success rate of all tested visualisation
methods, and we may thus recommend it as a first step in the analysis of
networked data.  However, if certain specific graph properties are to be
emphasized, then traditional graph drawing methods may be best instead
-- our experiments showed for instance that a Laplacian embedding is
well suited to making evident the diameter of a graph.  We must of
course stress that this method is only useful if general, structural
properties are to be summarised -- an exploration of the actual nodes
and edges of the network is, by construction, not possible with graph summarisation
methods.  The problem of visualising the individual nodes and edges of a
given large network is a separate problem that, in the end, cannot be
solved without showing only a small subset of the given network, giving
rise to the problem of letting the user search the nodes of a graph, for
instance using a search engine or even a recommender system, if the
nodes of the graph are amenable to that -- that problem is however
clearly outside of the scope of graph summarisation. 

Finally, we should address the question whether graph visualisation and
summarisation are a
necessary part of the graph mining process.  If certain specific
properties such as the bipartivity or assortativity of a given network
are to be detected, it is usually more productive to simply compute and
show the corresponding numerical statistics.  Graph visualisation and summarisation however
are usually used for a higher-level purpose:  As recently shown
\citep{dinosaur}, a general-purpose visualisation of a given dataset is
often a simple and effective way of starting an in-depth analysis.
Network datasets are of course not exempt from this, and thus a graph
summarisation method such
as SynGraphy can be recommended to fill that role, and, as we showed in
our experimental evaluation, give practitioners a better idea of a
given network dataset's overall structural properties.

An evident future extension of our method is to apply it to further
graph properties, although with more elaborate graph statistics, the
task of graph generation becomes more runtime-intensive, as the
vectorised expressions as given in Section~\ref{sec:graph-generation} are
then not possible anymore.

\bibliographystyle{spbasic}
\bibliography{ref,caricature,graph-generator,konect,kunegis} 

\end{document}